\pdfoutput=1
\documentclass[aps,pre,reprint,nofootinbib,superscriptaddress,longbibliography,floatfix]{revtex4-2}
\usepackage{array}[=2016-10-06]
\usepackage{amsmath,amssymb,mathtools}
\usepackage{graphicx,booktabs,microtype}
\usepackage{needspace}
\usepackage[hidelinks]{hyperref}
\newcolumntype{L}[1]{>{\raggedright\arraybackslash}p{#1}}
\newcommand{\E}{\mathbb{E}}

\newcommand{\KL}{D_{\mathrm{KL}}}
\newcommand{\TC}{\operatorname{TC}}
\makeatletter
\newenvironment{appendixtable}{%
  \par\addvspace{\baselineskip}\noindent
  \begin{minipage}{\linewidth}\def\@captype{table}%
}{\end{minipage}\par\addvspace{\baselineskip}}

\makeatother
\graphicspath{{figures/}}
\hypersetup{pdftitle={Local Predictability and Collective Fidelity in LLM-Agent Societies},pdfauthor={Igor Itkin}}

\newcommand{\QuestionGraphLL}{0.1518}
\newcommand{\QuestionGraphLLCI}{[0.1370, 0.1681]}
\newcommand{\QuestionGraphMAE}{0.0315}
\newcommand{\QuestionGraphMAECI}{[0.0188, 0.0465]}
\newcommand{\QuestionGraphW}{0.0526}
\newcommand{\QuestionGraphWCI}{[0.0392, 0.0694]}

\newcommand{\QuestionNeighbourLL}{0.0057}
\newcommand{\QuestionNeighbourLLCI}{[0.0031, 0.0084]}
\newcommand{\QuestionNeighbourMAE}{-0.0026}
\newcommand{\QuestionNeighbourMAECI}{[-0.0114, 0.0052]}

\newcommand{\QuestionHistoryLL}{0.0116}
\newcommand{\QuestionHistoryLLCI}{[0.0085, 0.0151]}
\newcommand{\QuestionHistoryMAE}{-0.0005}
\newcommand{\QuestionHistoryMAECI}{[-0.0066, 0.0049]}

\newcommand{\TransferGraphLL}{0.1412}
\newcommand{\TransferGraphLLCI}{[0.1216, 0.1635]}
\newcommand{\TransferGraphMAE}{0.0238}
\newcommand{\TransferGraphMAECI}{[-0.0059, 0.0556]}
\newcommand{\TransferGraphW}{0.0307}
\newcommand{\TransferGraphWCI}{[-0.0155, 0.0790]}

\newcommand{\TransferNeighbourLL}{0.0027}
\newcommand{\TransferNeighbourLLCI}{[-0.0082, 0.0150]}
\newcommand{\TransferNeighbourMAE}{-0.0253}
\newcommand{\TransferNeighbourMAECI}{[-0.0419, -0.0105]}
\newcommand{\TransferNeighbourW}{-0.0537}
\newcommand{\TransferNeighbourWCI}{[-0.0789, -0.0325]}
\newcommand{\TransferHistoryLL}{0.0103}
\newcommand{\TransferHistoryLLCI}{[0.0057, 0.0156]}
\newcommand{\TransferHistoryMAE}{0.0076}
\newcommand{\TransferHistoryMAECI}{[-0.0040, 0.0187]}

\newcommand{\MCMaxMAE}{0.0032}
\newcommand{\MCMaxW}{0.0058}

\newcommand{\FreshComplete}{430}
\newcommand{\FreshTotal}{440}
\newcommand{\FreshBalanced}{22}
\newcommand{\FreshTestEpisodes}{304}

\newcommand{\SemanticTransferGraphMAE}{0.0308\;[0.0038, 0.0608]}

\newcommand{\FreshQwenLocalLL}{0.0039\;[-0.0206, 0.0265]}
\newcommand{\FreshQwenLocalMAE}{-0.0101\;[-0.0343, 0.0121]}

\newcommand{\FreshGemmaLocalLL}{-0.0057\;[-0.0244, 0.0098]}
\newcommand{\FreshGemmaLocalMAE}{-0.0005\;[-0.0056, 0.0035]}

\newcommand{\SeparateQwenLocalLL}{0.0160\;[-0.0012, 0.0416]}
\newcommand{\SeparateQwenLocalMAE}{0.0016\;[-0.0193, 0.0264]}
\newcommand{\SeparateGemmaLocalLL}{0.0169\;[0.0023, 0.0350]}
\newcommand{\SeparateGemmaLocalMAE}{-0.0022\;[-0.0084, 0.0038]}

\newcommand{\LongOriginalComplete}{191}
\newcommand{\LongAmendedComplete}{192}
\newcommand{\LongOriginalPanels}{7}
\newcommand{\LongAmendedPanels}{8}
\newcommand{\LongQwenShortScalarLL}{0.2829}
\newcommand{\LongQwenShortScalarMAE}{0.1694}

\newcommand{\LongQwenShortHistoryLLGain}{0.0647\;[0.0210, 0.1115]}
\newcommand{\LongQwenShortHistoryMAEGain}{-0.0367\;[-0.0679, 0.0036]}
\newcommand{\LongQwenRefitScalarLL}{0.1687}
\newcommand{\LongQwenRefitScalarMAE}{0.2167}

\newcommand{\LongQwenScalarRefitLLGain}{0.1142\;[0.0747, 0.1632]}
\newcommand{\LongQwenScalarRefitMAEGain}{-0.0472\;[-0.0853, 0.0100]}
\newcommand{\LongGemmaShortScalarLL}{0.2629}
\newcommand{\LongGemmaShortScalarMAE}{0.0908}

\newcommand{\LongGemmaRefitScalarLL}{0.1880}
\newcommand{\LongGemmaRefitScalarMAE}{0.1000}

\newcommand{\TopicOriginalComplete}{478}
\newcommand{\TopicComplete}{479}
\newcommand{\TopicPanels}{39}

\newcommand{\TopicGemmaFrozenScalarMAE}{0.5467}

\newcommand{\TopicGemmaFrozenHistoryLLGain}{0.0200\;[0.0056, 0.0376]}
\newcommand{\TopicGemmaFrozenHistoryMAEGain}{-0.0149\;[-0.0240, -0.0043]}

\newcommand{\TopicGemmaAdaptScalarMAE}{0.2234}

\newcommand{\TopicGemmaScalarAdaptMAEGain}{0.3234\;[0.2014, 0.4691]}

\newcommand{\TopicQwenFrozenScalarMAE}{0.2806}

\newcommand{\TopicQwenFrozenHistoryLLGain}{0.0067\;[0.0029, 0.0104]}
\newcommand{\TopicQwenFrozenHistoryMAEGain}{0.0186\;[0.0140, 0.0227]}

\newcommand{\TopicQwenAdaptScalarMAE}{0.2100}

\newcommand{\TopicQwenScalarAdaptMAEGain}{0.0706\;[0.0058, 0.1418]}

\newcommand{\APIOriginalComplete}{119}
\newcommand{\APIComplete}{120}
\newcommand{\APIPanels}{40}

\newcommand{\InvGemmaLagAuto}{0.1307}
\newcommand{\InvQwenLagAuto}{0.0064}
\newcommand{\InvGemmaLagWarm}{0.0198}
\newcommand{\InvQwenLagWarm}{0.0049}
\newcommand{\InvGemmaWarmChangeMSE}{$+0.1758\;[+0.0280,+0.3927]$}
\newcommand{\InvGemmaWarmChangeCRPS}{$+0.0936\;[+0.0278,+0.1928]$}
\newcommand{\InvGemmaMarkovMSE}{0.0320}
\newcommand{\InvGemmaOffsetGraphMSE}{0.0276}

\newcommand{\InvGemmaMarkovCRPS}{0.0922}

\newcommand{\InvQwenMarkovMSE}{0.0166}

\newcommand{\InvQwenOffsetHistoryMSE}{0.1726}
\newcommand{\InvQwenMarkovCRPS}{0.0664}

\newcommand{\InvQwenOffsetHistoryCRPS}{0.2916}
\newcommand{\InvGemmaOffsetGain}{$+0.4713\;[+0.0952,+0.9418]$}
\newcommand{\InvGemmaDonorMSE}{0.3730}
\newcommand{\InvQwenClockGainMSE}{$+0.1421\;[+0.0492,+0.2435]$}
\newcommand{\InvQwenCappedGapMSE}{$+0.0286\;[+0.0075,+0.0567]$}
\newcommand{\InvQwenClockGainCRPS}{$+0.1189\;[+0.0551,+0.1733]$}
\newcommand{\InvQwenCappedGapCRPS}{$+0.0242\;[+0.0080,+0.0385]$}
\newcommand{\InvQwenClockBefore}{0.3306}
\newcommand{\InvQwenClockAfter}{0.1885}
\newcommand{\InvQwenClockPercent}{43.0}
\newcommand{\InvTextDifference}{$-10.5263\;[-17.7632,-3.9474]$}
\newcommand{\InvGemmaOwnTextDelta}{+0.00}
\newcommand{\InvGemmaCrossTextDelta}{+21.05}
\newcommand{\InvTextFlipGemma}{$+2.6316\;[-9.2105,+13.1579]$}
\newcommand{\InvTextFlipQwen}{$+2.6316\;[-1.9737,+7.8947]$}
\newcommand{\InvTextFlipDifference}{$+0.0000\;[-12.5000,+14.4737]$}

\newcommand{\ProsQwenHistoryMSEPercent}{11.8}
\newcommand{\ProsQwenBaselineMSERatio}{10.7}
\newcommand{\ProsQwenOffsetMSE}{0.14431}
\newcommand{\ProsQwenSimpleMSE}{0.01350}
\newcommand{\ProsQwenHistoryCRPSPercent}{7.3}
\newcommand{\ProsQwenBaselineCRPSRatio}{4.5}
\newcommand{\ProsQwenOffsetCRPS}{0.26070}
\newcommand{\ProsQwenSimpleCRPS}{0.05784}
\newcommand{\ProsGemmaBaselineMSELower}{0.00001372}

\newcommand{\BootPrecisionSeeds}{10}
\newcommand{\BootPrecisionGemmaUnresolved}{8}
\newcommand{\BootPrecisionGemmaMin}{-0.00016077}
\newcommand{\BootPrecisionGemmaMax}{+0.00011511}
\newcommand{\BootPrecisionQwenPositive}{3}

\begin{document}
\title{Local Predictability and Collective Fidelity in LLM-Agent Societies}
\author{Igor Itkin}
\email{ig.itkin@gmail.com}
\affiliation{Tel Aviv, Israel}
\date{September 2026}
\clubpenalty=10000
\widowpenalty=10000
\begin{abstract}
Compact surrogates could reduce the cost of simulating large language model
societies, but must reproduce collective behavior. We compare individual
predictions and collective forecasts using 9,455 published trajectories and
new experiments on opinion dynamics. Neighbor information improves individual
prediction in all 16 public-data settings and pooled collective forecasts on
held-out questions, although collective gains depend on transfer conditions.
Tests on 24 new statements do not confirm earlier contrasting history effects
in forecasts from the initial state. Qwen benefits from history after three
observed rounds. These findings motivate direct collective validation,
explicit limits on available observations, and comparisons with simple baselines.
\end{abstract}
\maketitle
\raggedbottom

\section{Introduction}
Predicting individual decisions and reproducing collective dynamics are
distinct modeling tasks. This distinction matters when replacing expensive
large language model (LLM) agents with statistical surrogates: a model fitted
to observed decisions must subsequently operate on interactions generated
by its own predictions. Errors can therefore propagate through the population
and alter its collective behavior. Here we investigate when improvements in
individual prediction translate into more accurate collective dynamics, using
opinion formation in LLM-agent societies as a test case.

Physics of Agents~\cite{physicsagents} provides a direct setting for evaluating
this question. El et al.\ fit statistical-mechanics models of opinion
updates on graphs whose links represent positive or negative relationships.
They evaluate individual predictions, simulated trajectories and collective
outcomes. Their public histories and code~\cite{dataset,code} let us compare
models with different inputs on the same test data. We ask whether inputs
that help predict an agent's next opinion also help predict the group's
behavior when the model generates all subsequent opinions itself.

Three distinctions motivate the tests. Knowing how many neighbors an agent
has may help differently from knowing their current opinions. Past opinions
may help predict the next recorded update without helping when the model
must generate its own history. Finally, fitting a model on a new question
or observing its first few rounds provides information unavailable to an
unchanged model started from the initial state. A model that predicts
individual updates more accurately need not predict the group's average
opinion more accurately. The probability bounds in
Eqs.~\eqref{eq:path} and~\eqref{eq:pinsker} make this distinction precise.

The framework of Poor Man's Agentic Modeling~\cite{poor} motivates our choices
of current and past information. This article is
a standalone empirical companion: it does not reproduce that preprint's
coarse-graining derivations or fitted-response experiments. Instead, its
reference is recorded LLM behavior, not behavior generated by the fitted
surrogate itself.

We report six trajectory studies (Table~\ref{tab:studies}). Within each study,
candidate models use the same test episodes. The public-data analysis compares
the population mean, neighbor counts, neighbor opinions and past opinions as
predictors, testing transfer to new questions and graphs. Further experiments
vary group size, forecast duration, questions and LLM backends. We then
investigate why adding history can change forecast quality. A final test on
24 additional statements fixes the predictors, adaptation rules and primary
comparisons before collecting the new responses.

In the public data, neighbor opinions improve individual prediction in all
reported point estimates and improve the pooled group forecast on held-out
questions. The group-level benefit changes with the transfer setting and
with the inclusion of question and persona features. Earlier tests found
opposite effects of history for the two local backends; the 24 new statements
do not confirm that contrast for forecasts started from the initial state.
For Qwen, history helps after three observed rounds. However, a simple
two-state model fitted to those rounds outperforms the graph model with an
adjusted intercept. These are results about forecasting procedures, not
evidence of intrinsic memory differences between LLM families.

We do not pool the studies into one effect: their questions, training data
and observed trajectories differ or overlap. Follow-up protocols were fixed
after earlier results had been inspected; they were not externally
preregistered. The
608 one-step text interventions remain exploratory
(Appendix~\ref{app:history-investigation}); the prospective 24-statement test
is reported separately (Sec.~\ref{sec:prospective}). Throughout, the simulator
tracks every agent, and we assess specified group forecasts rather than
a reduction of the population to a few variables.

\subsection{Related work}
\label{sec:related}
Experiments with interacting LLM agents have shown the emergence of shared
conventions and collective biases~\cite{ashery}. Several methods reduce the
cost of simulating such populations. Chopra et al.\ query an LLM for
representative groups of agents, then sample individual actions from the
resulting group-specific distributions. They test the resulting simulations
against infection and employment data~\cite{llmarchtypes}. Light Society
trains small neural networks to replace LLM decisions and checks both
individual responses and collective opinion trajectories. Its tests include
a 10{,}000-agent network with surrogate replacement ranging from 0\% to
100\%, including a fully LLM-driven reference
(Fig.~3i of Ref.~\cite{lightsociety}).

Other methods aim to reproduce observed human activity. MF-LLM uses a learned
summary of population history to guide individual decisions~\cite{mfllm}.
MF-MDP also tracks each agent's state and the distribution of states across
the population, and trains its forecasts over several future
steps~\cite{mfmdp}. Our reference is instead the recorded behavior of
LLM agents. Reproducing these simulations does not establish that they
accurately describe human societies.

Physics of Agents is the closest empirical study~\cite{physicsagents}.
It fits statistical-mechanics models to opinion updates and evaluates both
individual predictions and collective outcomes, including transfer to
unseen graphs. We reanalyse its public histories with different sets of
predictor inputs: the population mean, neighbor opinions and past opinions.
Our policies are not exact reproductions of its model, which also uses
information about questions and personas. We then test the same types of
predictors on separately collected LLM trajectories. Poor Man's Agentic
Modeling~\cite{poor} motivates these comparisons through its emphasis on
the information and history retained by a surrogate.

The difference between predicting recorded decisions and generating a
sequence of decisions is established in imitation learning~\cite{ross}:
an error can change the states encountered later. Projection methods also
show why omitting state variables can make past observations useful for
prediction~\cite{mori,zwanzig}. Neither observation identifies an intrinsic
memory mechanism in an LLM.

We use proper scoring rules to evaluate
probabilistic predictions~\cite{scoring} and directly measure errors in
the selected collective quantities. Our focus is whether adding neighbor
or past opinions improves those quantities, and whether the result changes
when a forecast starts from the initial state or after observed rounds.

\section{Study scope and information comparisons}
\label{sec:bridge}
We compare probabilistic models of individual opinion updates, called
\emph{policies}. Each policy is scored both on recorded LLM transitions and
on group trajectories that it generates. All simulators track agents
separately. The global-information policy uses the population mean alongside
each agent's own features. Graph policies also use neighbor counts and
opinions; history policies add values from previous rounds.
The label \emph{scalar} in the new-campaign tables refers to the global-mean
policy. Only the added population statistic is scalar, not the simulator's
whole state.

An \emph{autonomous} forecast starts from the initial state and generates
every later update itself. A \emph{conditional} forecast starts after a
specified sequence of observed rounds, called the observed prefix.

The public-data degree-only control tests the value of knowing neighbor
counts without knowing their opinions. It is also used in the first new
size experiment, but not in the longer, new-question or final API follow-ups.
In those follow-ups, the graph comparison adds counts and opinions together.
These tests do not establish the smallest sufficient set of inputs. Forecast
error may come from fitting, omitted information or sampling noise; we do not
uniquely separate those sources.

Table~\ref{tab:studies} distinguishes the six LLM-trajectory designs. Their
overlapping questions, calibration information and reference ensembles prevent
pooling them into one effect or an independent-replication count. The public-data
analysis is fixed at $N=32$; the new grid probes four sizes over three updates;
later studies emphasize horizons or questions. The API extension tests only
$N=16$, while the 24-statement replication returns to $N=16,64$ and separates
autonomous from observed-prefix forecasts. None is an asymptotic scaling
experiment.

\begin{table*}[t]
 \caption{Scope of the six LLM-trajectory studies. Episode counts include designated training
 and validation data, not only tests. Completed technical replacements count once;
 failed originals are preserved. $N$ and $T$ describe held-out evaluation.
 Counts and overlapping questions must not be interpreted as independent
 replications or pooled into a common effect.}
 \label{tab:studies}
 \centering\small
 \setlength{\tabcolsep}{4pt}
 \renewcommand{\arraystretch}{1.17}
 \begin{tabular}{@{}p{0.14\textwidth}p{0.12\textwidth}p{0.15\textwidth}p{0.53\textwidth}@{}}
 \toprule
 Study & Complete episodes & Test population and horizon & Held out and scope of inference\\
 \midrule
 Public histories & 9,455/9,600 & $N=32$; $T=8$ &
 18 held-out questions shared across four backbones. Two overlapping transfer
 settings: new questions, or new questions plus unsigned graphs. Reanalysis,
 not a prospective experiment or exact source-model reproduction.\\
 New finite grid & \FreshComplete/\FreshTotal & $N=16,32,64,128$; $T=3$ &
 New initializations and sizes, but calibrated questions. Five backends;
 balanced evaluation retains \FreshBalanced{} backend/question panels and
 \FreshTestEpisodes{} test episodes. Small or single-class panels limit inference.\\
 Long follow-up & \LongAmendedComplete/192 & $N=16,64$; $T=3,6,12$ &
 Five test initializations per size, four known subjective questions, two local
 backends. Short-frozen versus long-refit calibration; nested horizons are not
 independent tests. One transport-failed episode replaced.\\
 New statements & \TopicComplete/480 & $N=16,64$; $T=3,6$ &
 20 fixed new statements, two local backends. Untouched source policies versus
 topic-specific training/validation; two test initializations per size.
 Balanced panels retain 20 Gemma and 19 Qwen questions after one technical replay.\\
 API extension & \APIComplete/120 & $N=16$; $T=3,6$ &
 The same 20 new statements on Qwen Max and GLM, one test initialization each;
 both calibration arms. Sixteen separate source episodes precede the listed grid.
 One validation replay; no API population-size scaling test.\\
 Prospective replication & 192/192 & $N=16,64$; $T=6$ &
 24 fixed additional statements, two local backends and two test initializations
 per size. Primary scores use updates 4--6. Original source policies remain
 fixed; conditional forecasts observe states 0--3 with prespecified adaptation
 rules and simple baselines. No failures or exclusions (Sec.~\ref{sec:prospective}).\\
 \bottomrule
 \end{tabular}
\end{table*}

\section{Surrogate policies and evaluation targets}
\subsection{Retained information and policy features}
Let $s_i(t)\in\{-1,+1\}$ be a recorded binary opinion and
$J_{ij}\in\{-1,0,+1\}$ the symmetric, zero-diagonal communication graph.
Spin signs denote agreement/disagreement with the statement; edge signs denote
positive/negative relationships, with zero meaning no edge. With $\mathbf1$
the indicator of its subscripted event, define
\begin{align}
 m(t)&=N^{-1}\sum_i s_i(t),\\
 d_i^\pm&=\sum_j\mathbf{1}_{\{J_{ij}=\pm1\}},\\
 h_i^\pm(t)&=\sum_j\mathbf{1}_{\{J_{ij}=\pm1\}}s_j(t).
\end{align}
The mean opinion is $m(t)$; $d_i^\pm$ counts positive or negative neighbors,
and the corresponding field $h_i^\pm(t)$ sums their opinions.
Here $h_i^-$ sums opinions on negative edges without an additional minus sign;
its coefficient carries the fitted direction of influence. Using two fields removes the exact redundancy
of an additional unsigned field,
$h_i^{\mathrm{all}}=h_i^++h_i^-$, which is not included;
parameter identifiability still depends on the training design. Identifiability
here means uniqueness of coefficients from the observational distribution, not
merely a reproducible penalized optimum. The revised source paper also discusses
its coupling redundancy (Sec.~5.4 and Table~3 of Ref.~\cite{physicsagents}).
For example, at fixed degree $k$, $d_i^++d_i^-=k$ makes the
intercept and two degree coefficients linearly dependent.
Regularization selects a predictive fit but does not establish uniquely identified physical couplings.
Such a dependency does not by itself imply that the two field
coefficients are individually unidentifiable in every design.

Every fitted policy has the form
\begin{equation}
 \widehat p_i(t)=\sigma\!\left(b+\boldsymbol\theta^\top z_i(t)\right),
 \qquad \sigma(u)=(1+e^{-u})^{-1},
 \label{eq:policy}
\end{equation}
Here $\widehat p_i(t)$ is the conditional probability of the target next spin
$s_i(t+1)=+1$, $z_i(t)$ the train-standardized feature vector, $b$ the intercept,
and $\boldsymbol\theta$ the slope vector. Here $\sigma$ denotes the logistic function.
All policies share the common covariates $s_i(t)$, $s_i(0)$, and $t$.
The interaction-free policy uses only these covariates. The global policy adds
$m(t)$. A degree control also retains $d_i^+$ and $d_i^-$, but no neighbor
states. The graph policy adds $h_i^+$ and $h_i^-$ to that control. A normalized
neighbor policy instead uses $h_i^+/d_i^+$ and $h_i^-/d_i^-$, defining an empty
neighborhood mean as zero and retaining the degree covariates.

Two history policies augment the graph policy with
\begin{equation}
 \left\{s_i(t-r),h_i^+(t-r),h_i^-(t-r)\right\}_{r=1}^{L_{\mathrm{hist}}},
 \qquad L_{\mathrm{hist}}\in\{1,2\}.
 \label{eq:history-features}
\end{equation}
The lag $r$ counts rounds into the past; $L_{\mathrm{hist}}$ is the number of
past rounds retained. Unavailable negative-time values are padded with the initial state, both during
training and rollout. The policies have 3, 4, 6, 8, 8, 11, and 14 covariates,
respectively, plus an intercept. A persistence control always retains $s_i(0)$ in
autonomous simulation and predicts the current spin in one-step evaluation.

We vary the information supplied to the predictor while still simulating
every agent. The global policy does \emph{not} replace the population with
one variable; graph and history policies also retain individual opinions.
All these fitted families except
interaction-free retain $m(t)$, even when the local-input LLM prompt does not
supply it. This is a comparison of predictor information, not a reproduction
of each agent's prompt-level access constraints. During rollout, $m(t)$ is
computed from the surrogate's current spins, never observed future states.
Static degree controls isolate a possible
benefit of neighbor states from a benefit of knowing only how many neighbors
exist. None of the primary public-data policies uses ground-truth answers, question embeddings,
persona embeddings, or text messages. They are consequently not numerical
replications of the source authors' intrinsic-field model.

\subsection{Local scoring rules and collective fidelity}
Log loss evaluates probabilistic transitions, rather than only thresholded decisions;
its expected value is minimized by the true probability, making it a proper
scoring rule~\cite{scoring}. Nevertheless, lower log loss need not mean a
smaller error in a chosen group-level quantity. Accurate local transition
probabilities can still provide an upper bound on that error, as follows.

To make the distinction explicit, condition on a question, graph, and initial-state
law. Write $s(t)=(s_1(t),\ldots,s_N(t))$. Let $P$ be the true joint spin-path
law, $H_t=(s(0),\ldots,s(t))$, and $Q$ a
surrogate using independent synchronous Bernoulli draws conditional on its retained
history. Write $P_t=P(s(t+1)\mid H_t)$ and let $P_{it}$ be its one-node marginals.
Write $Q_{it}=Q(s_i(t+1)\mid H_t)$ for the surrogate one-node kernel evaluated
on that history. Relative entropy is
\begin{equation}
 \KL(P\|Q)=\int \log\!\left(\frac{dP}{dQ}\right)dP,
 \label{eq:kl-definition}
\end{equation}
with natural logarithms and value $+\infty$ if $P$ is not absolutely
continuous with respect to $Q$~\cite{coverthomas}. For any observed history,
\begin{equation}
 \KL\!\left(P_t\middle\|\prod_i Q_{it}\right)
 =\TC(P_t)+\sum_i\KL(P_{it}\|Q_{it}),
 \label{eq:tc}
\end{equation}
where $\TC(P_t)=\KL(P_t\|\prod_iP_{it})$ is total correlation~\cite{watanabe} of the next-spin distribution at that
fixed history; subsequent expectations average over histories.
This follows by adding and subtracting $\sum_i\log P_{it}$ inside the expectation.
Summing Eq.~\eqref{eq:tc} over time and taking expectation under $P$ gives the
conditional chain rule for path divergence. With identical initial-state laws,
\begin{equation}
 \KL(P\|Q)=\sum_{t=0}^{T-1}\E_P\!
 \left[\TC(P_t)+\sum_i\KL(P_{it}\|Q_{it})\right].
 \label{eq:path}
\end{equation}
For a bounded path observable $O$, let
$\operatorname{osc}(O)=\sup_\omega O(\omega)-\inf_\omega O(\omega)$,
where $\omega$ ranges over spin paths. Pinsker's inequality~\cite{coverthomas} then gives
\begin{equation}
 |\E_P O-\E_Q O|
 \leq\operatorname{osc}(O)\sqrt{\KL(P\|Q)/2}.
 \label{eq:pinsker}
\end{equation}
These are standard probability identities, included to specify the claim precisely.

Lower expected one-step cross entropy, averaged over the true distribution of
complete observed histories with the same weighting, decreases path divergence
for this product-kernel family. It does not guarantee that the error of each
particular $O$ decreases: a smaller upper bound in Eq.~\eqref{eq:pinsker} need
not mean a smaller actual error. Finite test
data estimate cross entropy, not the unknown entropy or conditional total correlation.
The $NT$ accumulation also matters. Consequently, a small per-node improvement
cannot alone certify a small absolute path divergence. Direct autonomous evaluation
remains necessary for the observables at issue here.

History features should be interpreted with similar care. Information discarded
when messages are projected to spins can induce effective memory in the projected
process, consistent with the general logic of projection methods~\cite{mori,zwanzig}.
A predictive lag is not by itself a measured memory kernel, proof of causal memory,
or evidence that a finite set of spin lags closes the dynamics.

\section{Data, protocol, and evaluation}
\subsection{Snapshot and integrity audit}
We use a pinned snapshot of the public \texttt{agent-opinions} Parquet dataset~\cite{dataset};
Appendix~\ref{app:audit} describes the audit and points to the deposited revision
and file hash.
The snapshot and original analysis protocol were fixed on 6 September 2026.
Physics of Agents v2 appeared on 9 September and also reports a 64-agent
mixed-model experiment (Appendix~E.6 of Ref.~\cite{physicsagents}). Our public-data
results below use the fixed 32-agent snapshot, not those additional runs.
Our file-level audit finds 9,600 episodes, 60 distinct questions, four backbones,
$N=32$ nodes, and nine recorded states ($t=0,\ldots,8$) per episode. There are 40
objective and 20 subjective questions. The backbones are GPT-4o-mini, Qwen3.5-9B,
Gemma-3n-E4B-it, and Llama-3.1-8B-Instruct. The full union contains 14 distinct
adjacency matrices; a question does not necessarily appear on all 14.

The audit validates shapes, node/time alignment, symmetric signed adjacency with
zero diagonal, binary recorded histories, five raw votes per node-time, and agreement
between each recorded state and the sign of the raw-vote sum. Duplicate episode
identifiers are rejected. Of 9,600 episodes, 38 have nonbinary recorded histories and
107 additional episodes disagree with the raw majority. The primary analysis retains
9,455 complete episodes. Raw zero votes are retained only when the resulting majority
is unambiguous and agrees with the recorded spin. They are never recoded as $-1$.
Appendix~\ref{app:audit} gives the backbone-specific exclusions and their limitations.

The graph union contains twelve signed graphs and two unsigned lattice graphs.
The latter have 52 and 73 undirected edges and nonuniform degrees because of their
boundaries. We identify the unsigned graphs using the nonnegative-adjacency convention
in the pinned analysis code~\cite{code}; no graph is inferred from node ordering alone.

Our observables concern the binary spin $s_i$, not the mean of the five raw votes.
In particular, $N^{-1}\sum_i s_i^2=1$ identically. We therefore do not use this quantity
as conviction or claim to reproduce the source study's continuous-opinion distinction
between indifference and polarization.
The public target is the sign of a five-vote sum, whereas a new-campaign spin
is the opinion in a single response (Sec.~\ref{sec:fresh}). These observation rules
define different transition targets. Cross-study differences therefore cannot
be attributed to backend identity alone.

\subsection{Question-level separation and two transfer settings}
Throughout this article, \emph{calibration} means fitting the small surrogate's
response probabilities. It neither updates the LLM's weights nor corrects its
held-out answers. Training data determine surrogate coefficients and feature
scaling; validation data select regularization (and, where specified, history
depth); test data only measure performance. A test on a new trajectory of a
calibrated question checks a different generalization claim from a test on a
question never used in fitting. We report that distinction for every follow-up.
These separations guard against test leakage; they do not prove that finite-data
overfitting or domain shift is absent.

Questions are hashed from their mode and statement, sorted, and shuffled with a fixed
seed separately within each mode. The 50/20/30 split yields 20/8/12 objective and
10/4/6 subjective questions for training/validation/test. A question has the same
assignment across all backbones, graphs, and replicas. No question enters more than
one partition.

In \emph{question transfer}, available graph types are retained in all partitions.
In \emph{question-plus-graph transfer}, training and validation use only signed
graphs, while test evaluation uses only the two unsigned lattices on held-out
questions. Thus the latter changes the graph distribution and questions together;
it is not a controlled topology-only intervention. It also changes the edge-sign
distribution and degree profiles. The two settings contain 2,841 and 571 accepted
test episodes, respectively. The second is a subset of the first test population,
not an independent dataset.

Each backbone and task mode is fitted separately. Covariates are standardized using
training means and population standard deviations only; a standard deviation
below $10^{-10}$ is replaced by one. We minimize
\begin{equation}
 \frac1n\sum_{a=1}^{n}\left[\log(1+e^{\eta_a})-y_a\eta_a\right]
 +\frac{\|\boldsymbol\theta\|_2^2}{2Cn},
 \label{eq:fit}
\end{equation}
where $a$ indexes the $n$ training node transitions, $s_a$ is the target next
spin, $y_a=(s_a+1)/2$, and $\eta_a=b+\boldsymbol\theta^\top z_a$. The intercept
is unpenalized and $C>0$ sets the inverse penalty strength in this normalization.
For each policy,
$C\in\{0.01,0.1,1,10,100\}$ is chosen by mean validation log loss, weighting validation
questions equally. There is no refitting on validation data. History depth is likewise
selected by validation log loss between the one- and two-lag policies; two lags are
selected in every cell. All seven policies are evaluated regardless of their test
ranking. A cell denotes a backbone, task mode, and transfer setting, giving 16 cells.

The data audit, protocol, source hashes, and question split were frozen locally before
the first fit on 6 September 2026. The source paper and data schema were already known.
This is not an external preregistration or a blind prospective test. Scientific
predictions are evaluated as outcomes, not encoded as software tests that must pass.

\subsection{Observed transitions and autonomous collective observables}
For a target $y\in\{0,1\}$ and predicted probability $p$, the individual
log loss is $-y\log p-(1-y)\log(1-p)$ and the Brier loss is $(y-p)^2$~\cite{scoring}.
Probabilities are clipped to $[10^{-12},1-10^{-12}]$ for numerical log loss;
this does not modify the population identities above. We denote the aggregated
log loss by $\ell$, measured in nats per node transition.
One-step log loss is first averaged over nodes and rounds within an episode, then
episodes within a question, then test questions. Brier score and the source-style
four-way flip/stay-by-next-sign balanced accuracy~\cite{physicsagents} are secondary diagnostics.
The latter averages classification accuracy equally over the four strata
(flip or stay) $\times$ (next sign $+1$ or $-1$), using $p\ge1/2$ to predict $+1$. A missing
stratum would be flagged rather than assigned an invented score. Persistence probabilities
are clipped to $[10^{-12},1-10^{-12}]$ for log loss; its score therefore depends on this
convention and is not a calibrated probabilistic baseline.

For collective evaluation, every accepted test initial state seeds $R=200$ autonomous
trajectories. At each round, the surrogate recomputes its own features and draws all
next spins independently and synchronously using Eq.~\eqref{eq:policy}. No observed
future state or future lag is fed back. Candidate policies share random-number
streams for each initial state. Four observed replicas are nominally available per
question/graph configuration; two to four remain in the accepted test configurations.

We call the population's mean binary opinion its \emph{magnetization}.
For configuration $c=(q,g)$, where $q$ identifies a question and $g$ a graph,
let $K_c$ be its accepted replica count, $k$ index observed replicas and $r$
index surrogate draws. Let $m_{ck}(t)$ be the
observed magnetization, and $\widehat m_{ckr}(t)$ the simulated one. Define the
mean absolute error (MAE) of the ensemble-mean trajectory by
\begin{equation}
 E_m(c)=\frac1T\sum_{t=1}^{T}
 \left|\frac1{K_c}\sum_km_{ck}(t)
 -\frac1{K_cR}\sum_{k,r}\widehat m_{ckr}(t)\right|.
 \label{eq:mae}
\end{equation}
Ensembles are averaged \emph{before} taking the absolute difference; the trivially
matched initial time is excluded. The analogous score for $|m(t)|$ and the
Wasserstein-1 distance between observed and predicted terminal magnetizations are
secondary collective outcomes. The $|m|$ score replaces each trajectory
magnetization in Eq.~\eqref{eq:mae} by its absolute value before ensemble
averaging. For terminal empirical distributions with cumulative distribution
functions $F$ and $G$, $W_1=\int_{-1}^{1}|F(x)-G(x)|\,dx$~\cite{optimaltransport}.
The latter uses the $K_c$ observed endpoints and
$K_cR$ simulated endpoints with equal empirical weights. Scores are averaged equally
over available graphs within a question, then over questions.

Paired reductions are defined as baseline minus candidate error, so positive values
favor the candidate. We use 2,000 paired percentile bootstrap resamples~\cite{efron} of whole
questions, each retained with all its associated measurements.
For pooled estimates, backbone-specific paired differences are averaged within each
question \emph{before} resampling the 18 test questions. Backbones are fixed repeated
measurements, not 72 independent questions; agents and rounds are not independent
uncertainty units. The pooled weighting gives the twelve objective questions twice
the total weight of the six subjective questions. Intervals are descriptive 95\%
cluster-bootstrap intervals without familywise multiplicity adjustment.
Throughout, a contrast is called \emph{resolved} when its stated interval
excludes zero, under the nominal or adjusted convention specified for that
comparison. This does not by itself establish practical importance.

\section{Public-data results}
\subsection{Graph information and collective prediction}
\begin{figure*}[t]
 \centering\includegraphics[width=\textwidth]{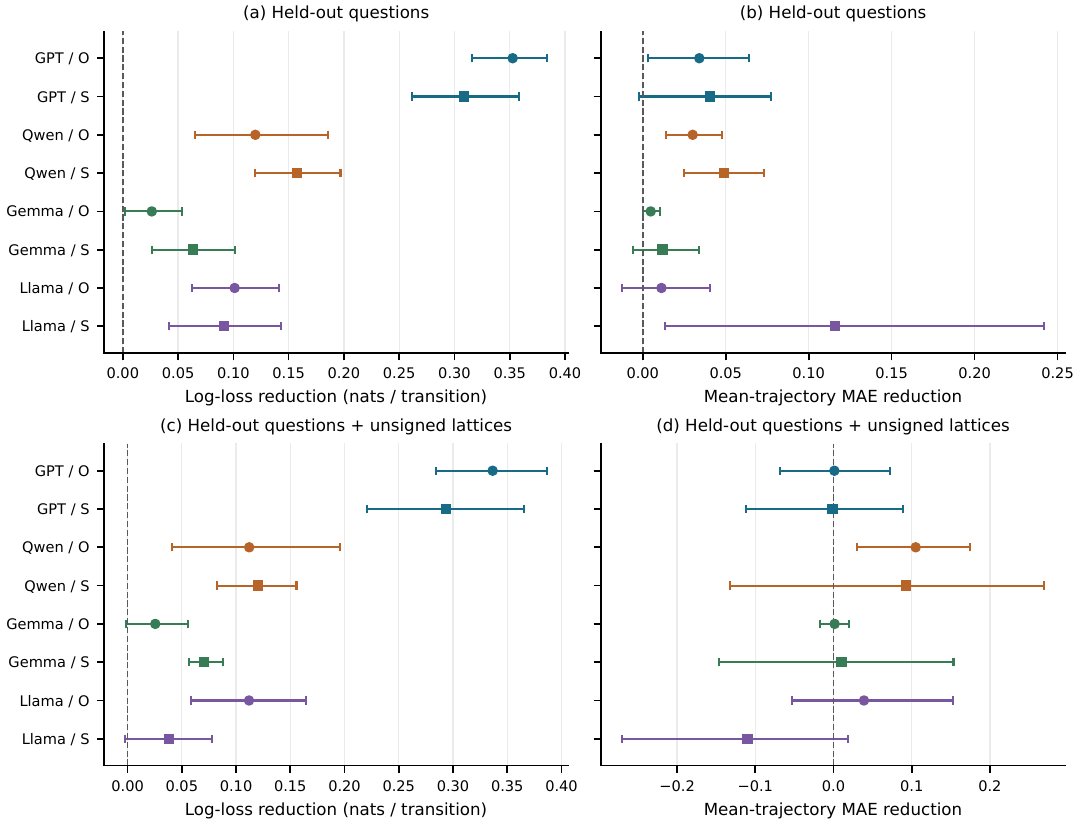}
 \caption{Graph-field policy versus global-mean policy on the same test questions.
 Positive values indicate lower error with graph fields. Points are paired reductions;
 bars are 95\% question-cluster bootstrap intervals. O and S denote objective and
 subjective questions (12 and 6 test clusters per backbone). Upper panels retain
 available graph types; lower panels train on signed graphs and test on unsigned
 lattices as well as new questions. Local log-loss gains are more consistent than
 collective-error gains. Intervals are not multiplicity-adjusted.}
 \label{fig:graph}
\end{figure*}
Graph fields lower one-step log loss relative to the global policy in all 16 cells
(Fig.~\ref{fig:graph}); 14 cell-level intervals exclude zero on the positive side.
The result also holds against the static degree control in all 16 point estimates,
with 15 positive intervals. Thus the measured local predictive benefit is not
explained simply by giving the policy the two node degrees.

For held-out questions, the pooled graph-versus-global reduction is
$\Delta\ell=\QuestionGraphLL$ nats per node-transition, with interval
$\QuestionGraphLLCI$. The collective reduction is
$\Delta E_m=\QuestionGraphMAE$ with interval $\QuestionGraphMAECI$.
Terminal-distribution error also improves, by $\QuestionGraphW$ with interval
$\QuestionGraphWCI$. Both the local and collective comparisons therefore favor
retaining neighbor states on average in this setting.

The simultaneous question-and-graph shift preserves a strong local benefit:
$\Delta\ell=\TransferGraphLL$ with interval $\TransferGraphLLCI$.
The collective estimate is $\Delta E_m=\TransferGraphMAE$, but its interval
$\TransferGraphMAECI$ crosses zero. The terminal $W_1$ reduction,
$\TransferGraphW$ with interval $\TransferGraphWCI$, is similarly unresolved.
These intervals do not establish absence of a benefit; the average collective-error
reduction is unresolved on these held-out questions under the joint shift.

Across the sixteen cells, primary-run mean-trajectory error improves in 14 point
estimates and terminal $W_1$ improves in 13. These descriptive counts should not be
treated as independent binomial trials: cells share questions, and tiny differences
can change sign with Monte Carlo sampling. Only six cell-level intervals exclude
zero positively for either of these collective comparisons. Table~\ref{tab:pooled}
reports all planned pooled information ablations, and Table~\ref{tab:cells} gives
absolute scores for the main policies.

\begin{table*}[t]
 \caption{Pooled paired reductions with descriptive 95\% intervals. Each question is
 averaged across the four backbones before bootstrap resampling; $n=18$ questions
 per setting. Positive numbers favor the second policy. $\Delta\ell$ is in nats
 per node-transition; $\Delta E_m$ is in binary-magnetization units.}
 \label{tab:pooled}
 \centering\small
\begin{tabular}{llrr}
\toprule
Setting & Comparison & $\Delta\ell$ & $\Delta E_m$\\
\midrule
Questions & Global to graph & $0.1518\;[0.1370, 0.1681]$ & $0.0315\;[0.0188, 0.0465]$\\
Questions & Degree to graph & $0.1515\;[0.1361, 0.1683]$ & $0.0318\;[0.0188, 0.0478]$\\
Questions & Graph to normalized & $0.0057\;[0.0031, 0.0084]$ & $-0.0026\;[-0.0114, 0.0052]$\\
Questions & Graph to history & $0.0116\;[0.0085, 0.0151]$ & $-0.0005\;[-0.0066, 0.0049]$\\
\midrule
Questions + lattices & Global to graph & $0.1412\;[0.1216, 0.1635]$ & $0.0238\;[-0.0059, 0.0556]$\\
Questions + lattices & Degree to graph & $0.1424\;[0.1224, 0.1656]$ & $0.0363\;[-0.0013, 0.0758]$\\
Questions + lattices & Graph to normalized & $0.0027\;[-0.0082, 0.0150]$ & $-0.0253\;[-0.0419, -0.0105]$\\
Questions + lattices & Graph to history & $0.0103\;[0.0057, 0.0156]$ & $0.0076\;[-0.0040, 0.0187]$\\
\bottomrule
\end{tabular}

\end{table*}

\subsection{Degree normalization and lagged features}
Normalized neighborhood means improve one-step log loss relative to unnormalized
graph sums in 13 of the 16 cell-level point estimates. On held-out questions the
pooled reduction is $\QuestionNeighbourLL$ with interval $\QuestionNeighbourLLCI$.
The corresponding trajectory-error reduction is $\QuestionNeighbourMAE$ with
interval $\QuestionNeighbourMAECI$: a local gain without a resolved collective gain.
Under the combined graph/question shift, the local difference is unresolved
($\TransferNeighbourLL$, interval $\TransferNeighbourLLCI$), while the collective
difference favors the \emph{unnormalized} graph fields: the reduction from graph
to normalized policy is $\TransferNeighbourMAE$, interval $\TransferNeighbourMAECI$.
Terminal $W_1$ likewise worsens under normalization, with reduction
$\TransferNeighbourW$, interval $\TransferNeighbourWCI$.

\begin{figure*}[t]
 \centering\includegraphics[width=\textwidth]{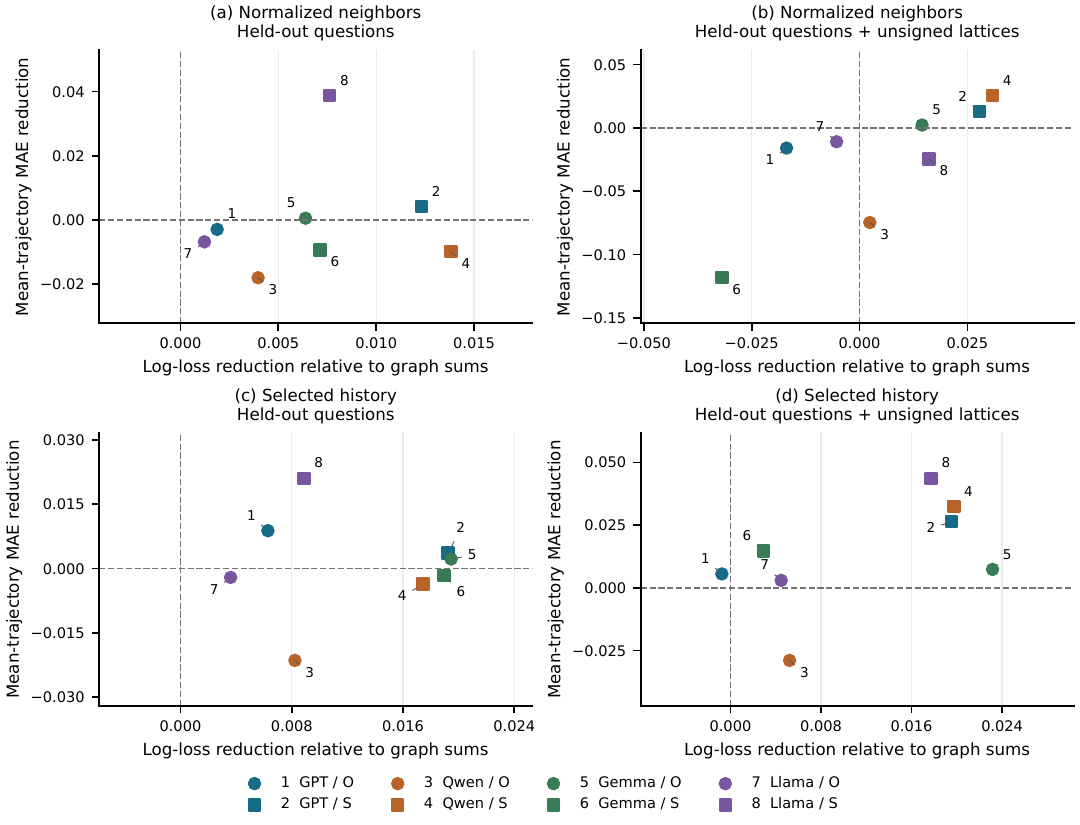}
 \caption{Local and collective changes relative to the graph-sum policy. Top:
 replace graph sums by degree-normalized neighborhood means. Bottom: add the
 validation-selected history. Rightward and upward changes improve local and
 collective scores, respectively; a lower-right point improves local prediction
 but worsens the collective observable. Colors identify backbones and point labels identify legend entries; circle/square
 markers denote objective/subjective questions, with backbone and mode identified
 in the legend. All sixteen cells appear for each comparison. These are point estimates;
 pooled intervals appear in Table~\ref{tab:pooled}.}
 \label{fig:ablations}
\end{figure*}

Normalization removes the multiplicative degree dependence of a field while keeping
degree as an additive covariate. These policies are not equivalent coordinate
systems for a heterogeneous-degree logistic response. Their different behavior is
therefore a response-model comparison, not proof that degree normalization is
generally inappropriate. Figure~\ref{fig:ablations} shows every cell, including
instances where local and collective rankings agree.

The validation-selected history policy lowers test log loss in 15 of 16 cells.
The pooled improvement over current graph fields is $\QuestionHistoryLL$
($\QuestionHistoryLLCI$) for question transfer and $\TransferHistoryLL$
($\TransferHistoryLLCI$) for combined transfer. Both intervals are positive.
Yet the pooled mean-trajectory reduction is $\QuestionHistoryMAE$
($\QuestionHistoryMAECI$) in the first setting and $\TransferHistoryMAE$
($\TransferHistoryMAECI$) in the second. Neither interval excludes zero.
Terminal-distribution differences are also unresolved in both settings.

These findings support the predictive relevance of earlier spin configurations
within this restricted model family. They do not establish a finite-memory closure
or show that history is useless for collective simulation. Rather, optimizing
one-step history depth does not deliver a demonstrated improvement in the selected
collective outcomes here.

\subsection{Absolute fidelity, simple controls, and Monte Carlo sensitivity}
Relative gains should not obscure substantial absolute error. For example, Qwen
objective-question graph-policy trajectory MAE is approximately 0.49 in both settings, despite
large local improvements. In Llama subjective-question combined transfer, the global
and graph errors are 0.421 and 0.531, respectively, while graph log loss improves
from 0.183 to 0.145. The interaction-free policy has trajectory MAE 0.171 in this
cell, despite worse log loss (0.224). These are different evaluation targets,
not a contradiction.

\begin{figure*}[t]
 \centering\includegraphics[width=\textwidth]{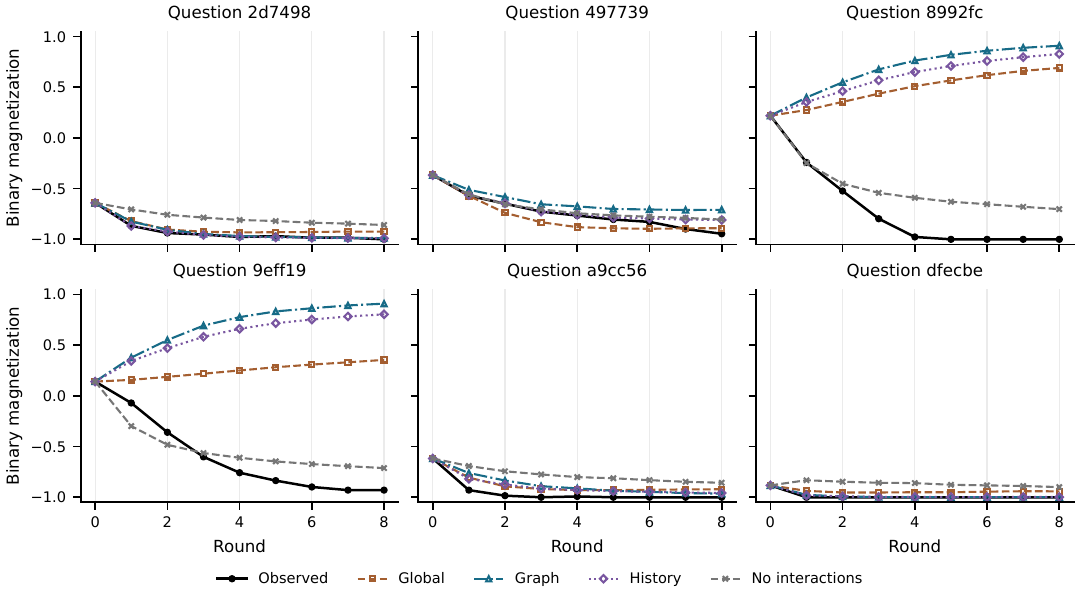}
 \caption{A post-hoc diagnostic of the Llama subjective-question combined-transfer
 cell, showing all six test questions rather than selected individual successes or
 failures. Black curves are observed binary magnetizations; colored curves are
 autonomous surrogate means from the same accepted initial states. All available
 unsigned graphs are averaged equally for each question. Question labels are fixed
 hash prefixes. This illustration has only two to four observed replicas per graph
 and should not be interpreted as an exact reference expectation.}
 \label{fig:trajectories}
\end{figure*}

For a descriptive view, Fig.~\ref{fig:trajectories} shows \emph{all six} held-out
subjective questions for this Llama cell. The cell was chosen after inspecting the
results because it illustrates a clear reversal; it is not an unbiased sample of
the sixteen cells and is not used for model selection. The displayed curves average
graphs after averaging replicas and simulator draws. The formal score in
Eq.~\eqref{eq:mae} takes absolute differences within each graph first, so error must
not be read directly from the graph-averaged illustration.

Persistence is a useful deterministic control but a poor calibrated transition
predictor. The graph policy beats it in mean-trajectory error in 11 of the 16 cells.
Its pooled improvements over persistence are positive in both settings, although
some individual systems are adequately described by little collective motion and
can favor simpler controls. Agreement of a mean trajectory alone therefore does
not identify the microscopic interaction mechanism.

After the primary analysis, we repeated the global, graph, and selected-history
rollouts with $R=1000$ and an independent seed, retaining exactly the fitted weights
and test populations. Across the 48 policy--cell combinations, the maximum absolute
change in reported trajectory MAE was $\MCMaxMAE$, and in terminal $W_1$ it was
$\MCMaxW$. The pooled graph-versus-global trajectory reductions became 0.0315 and
0.0240 for question and combined transfer, respectively, with the same qualitative
interval conclusions. The pooled history reductions also retained their conclusions.
This checks simulator Monte Carlo error, not uncertainty from the small observed
LLM ensemble, the question split, or model specification.

\subsection{Semantic, integrity, and split sensitivity}
Three post-primary extensions leave the original estimates intact. First, we add
the same 15 static semantic covariates to every fitted policy: three persona
principal components, three question components, and their nine products. Question
principal component analysis (PCA) is fitted only on training questions, separately by task mode; persona PCA uses
the fixed 32-persona upstream roster without outcomes. Regularization and history
selection still use only validation log loss. This is a matched augmentation,
not the source authors' two-stage estimator. The input vectors are the
pinned upstream \texttt{persona\_embeddings.json} and
\texttt{question\_embeddings.json} in the objective/subjective analysis folders~\cite{code};
we center the vectors and use the leading three right singular vectors.
The saved transforms and node ordering are retained with the sensitivity archive.

The semantic graph-versus-global comparisons are shown in Table~\ref{tab:semantic}.
Under joint question/graph shift the trajectory reduction becomes
$\SemanticTransferGraphMAE$: its interval now excludes zero. The primary unresolved
interval is therefore not evidence that graph information lacks collective value
under shift. The semantic history comparison remains locally positive but has an
unresolved mean-trajectory effect in both settings. Both the original and
semantically augmented fits therefore improve local history prediction without
a resolved mean-trajectory gain.

\begin{table*}[t]
 \caption{Post-primary matched semantic augmentation. Paired reductions and
 descriptive 95\% question-cluster intervals; positive favors the second policy.
 Each setting uses the original 18 held-out questions and accepted histories.}
 \label{tab:semantic}\centering\small
\begin{tabular}{llrr}
\toprule
Setting & Semantic comparison & $\Delta\ell$ & $\Delta E_m$\\
\midrule
Questions & Global to graph & $0.1510\;[0.1361, 0.1677]$ & $0.0309\;[0.0171, 0.0464]$\\
Questions & Graph to history & $0.0112\;[0.0082, 0.0145]$ & $-0.0022\;[-0.0092, 0.0037]$\\
Questions + lattices & Global to graph & $0.1420\;[0.1225, 0.1643]$ & $0.0308\;[0.0038, 0.0608]$\\
Questions + lattices & Graph to history & $0.0097\;[0.0054, 0.0148]$ & $0.0045\;[-0.0066, 0.0148]$\\
\bottomrule
\end{tabular}

\end{table*}

Second, five additional question-split seeds produce positive pooled graph log-loss
intervals in both transfer settings. The collective interval includes zero for
three of five question-transfer splits and all five joint-transfer splits.
These overlapping splits are sensitivity analyses, not independent replications.
Third, accepting any binary recorded history without requiring raw-majority
agreement retains 9,562 episodes; Table~\ref{tab:robustness} reports the changed
pooled estimates, and the archive retains comparisons on the common strict-reference
test episodes. Reconstructing a binary
majority retains exactly the same 9,455 histories as the primary filter, with no
changed node-times, so it is not an independent robustness result. The 112 additional
split/integrity cells and 16 semantic cells are complete; Appendix~\ref{app:extensions}
reports the full pooled split/integrity estimates. None of these checks identifies
the unobserved valid outcomes of episodes with nonbinary recorded states.

\section{Population-size transfer with calibrated questions}
\label{sec:fresh}
\subsection{Design, calibration, and incomplete outcomes}
We separately generated new trajectories on five installed or API backends, at
$N=16,32,64,128$ and $T=3$ synchronous updates. The two local backends use eight
questions each (four objective and four subjective); Qwen Flash and DeepSeek Flash
use four, and GLM uses two. These are not the public dataset's model snapshots.
Model identifiers, decoding settings and provenance are given in
Appendix~\ref{app:fresh}. Each population starts with shuffled, exactly balanced
spins and scripted stance-only messages on a randomly relabeled degree-four signed
ring: before relabeling, node $i$ is linked to $i\pm1$ and $i\pm2$ modulo $N$,
with each undirected edge independently attractive with probability 0.75.
Appendix~\ref{app:protocol-inputs} gives the exact prompts, persona roster and question lists. Initial graph and state
are paired across backends and information regimes for each question/size/replica.

Every agent receives a fixed persona and its own previous spin and public message.
The global-input regime adds the population mean; the local-input regime instead
adds signed neighbors' spins and public messages. This changes semantic content
as well as information topology. There is no explicit multi-round prompt history,
but the previous message can carry information beyond the binary spin.

The collection plan was locally frozen before generation. Training uses replica 0
at $N=16$, and regularization selection uses replica 0 at $N=32$. Test replicas are
1 and 2 for local backends, and 1 for API backends, at every size. Questions overlap
calibration and test: this tests size and initial-condition transfer, not transfer
to unseen questions. We retain persistence and the global-information (``scalar''), degree, graph, one-lag,
and two-lag families, with four matched persona indicators for every fitted policy.
The scalar family is still node-resolved, not a scalar macrostate closure.

Before fitting, an analysis clarification interpreted the plan's ``per model and
question'' unit literally: pool the two regimes for each fitted policy, without a
regime indicator. This imposes a common response law across regimes and may be
misspecified. Completion counts, but not fitted scores, were known at clarification;
this step is not retrospective preregistration. Section~\ref{sec:separate} separately
checks regime-specific calibration after inspecting the primary results.

All \FreshTotal{} planned episodes were processed, of which \FreshComplete{} are
complete. Eight Qwen Flash and two DeepSeek episodes fail the frozen binary-output
schema; all ten are in the global-input regime. No failed vote was repaired or
retried, and incomplete rounds do not become partial population states. Missingness
therefore cannot be treated as random. Requiring complete calibration in both regimes
permits 23 of 26 backend/question pairs. Requiring every planned test size, replica
and regime leaves \FreshBalanced{} pairs and \FreshTestEpisodes{}
test episodes. The primary balanced panel has eight questions for each local backend,
two objective questions for Qwen Flash, two subjective questions for DeepSeek, and
one objective plus one subjective question for GLM. The API panels are not comparable
samples of question types. Seven complete test episodes from the remaining calibratable
DeepSeek question are saved as available-case diagnostics, not substituted into the
balanced curves.

For each complete test episode $e$, $R=1000$ autonomous surrogate draws estimate
the mean magnetization. For a scored horizon $T$, define the single-reference-path error
\begin{equation}
 E_m^{\mathrm{new}}(e)=\frac{1}{T}\sum_{t=1}^{T}
 \left|m_e(t)-\frac{1}{R}\sum_{r=1}^R\widehat m_{er}(t)\right|.
 \label{eq:mae-new}
\end{equation}
Here $T=3$; the later campaigns use their stated scored horizons. Their tables
abbreviate $E_m^{\mathrm{new}}$ as $E_m$ and explicitly refer to
Eq.~\eqref{eq:mae-new}. Unlike Eq.~\eqref{eq:mae}, the observed reference is one path per initialized graph;
errors are subsequently averaged over replicas, then equally over questions and
sizes. This noisy-reference error is not an unbiased estimate of the discrepancy
between two expected magnetizations. We save per-round Monte Carlo standard errors
and per-replica scores. Question-cluster intervals use 2,000 resamples; no inferential
interval is reported for panels with fewer than four questions. Undefined binary
or four-transition balanced-accuracy categories are retained as undefined.

There is also a target--score distinction: absolute loss is consistent for a
conditional median, not a conditional mean~\cite{pointforecast}. Consequently,
Eq.~\eqref{eq:mae-new} can prefer a misspecified predictive law even to the true
law when each is represented by its mean. It remains the frozen descriptive
outcome, but it cannot alone establish fidelity of expected magnetization or its
distribution. Section~\ref{sec:proper-scores} adds a post-hoc sensitivity using
mean squared error and a proper distributional score on the same test panels.

\subsection{Finite-grid outcomes and calibration dependence}
\label{sec:separate}
\begin{figure*}[t]
 \centering\includegraphics[width=\textwidth]{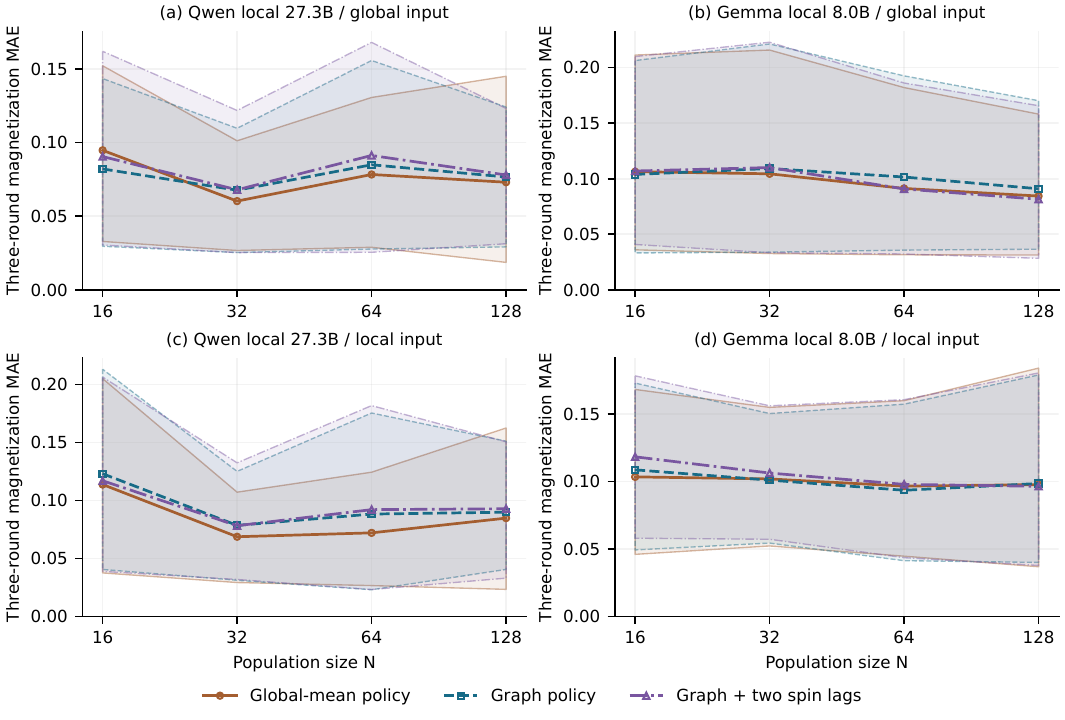}
 \caption{New LLM trajectories: balanced eight-question panels for the two local
 backends, with two test initializations per question/size/regime. Curves show
 three-round error against observed magnetization for the primary pooled-regime
 fits; shading gives descriptive 95\% question-cluster intervals. The same questions
 enter every size. Larger populations do not uniformly resolve policy error. Bands
 overlap and are not tests of paired differences; paired contrasts are reported
 in Table~\ref{tab:fresh}. These four sizes do not establish asymptotic scaling. $E_m$ abbreviates $E_m^{\mathrm{new}}$ in Eq.~\eqref{eq:mae-new}.}
 \label{fig:fresh}
\end{figure*}

Figure~\ref{fig:fresh} shows the fully observed local backends, and
Table~\ref{tab:fresh} gives all five backend panels. The primary local-input
graph-versus-scalar comparison is not a resolved advantage for either local backend:
Qwen has $\Delta\ell=\FreshQwenLocalLL$ and
$\Delta E_m^{\mathrm{new}}=\FreshQwenLocalMAE$; Gemma has
$\Delta\ell=\FreshGemmaLocalLL$ and
$\Delta E_m^{\mathrm{new}}=\FreshGemmaLocalMAE$.
The curves do not establish a universal decrease with $N$, let alone a power-law
exponent or a nonzero asymptotic floor.

Nine of the 23 calibratable backend/question pairs have a single next-spin class
throughout training, including all four objective questions for local Qwen.
For these, the frozen fitter uses the constant probability
$\widehat p=\frac{n_++1/2}{n+1}$, where $n_+$ is the number of positive
training targets and $n$ the training count; all slope coefficients are zero
and all fitted families coincide. This limits the campaign's ability to
distinguish models with different inputs. If all training responses agree,
a short consensus experiment may say little about which interaction information
is needed. The API results are exploratory: for example, pooled-regime
DeepSeek local input gives a positive local point difference but a negative collective
one, on only two subjective questions.

\begin{table*}[t]
 \caption{New-campaign balanced-panel graph-versus-scalar reductions, averaged
 equally across the four sizes. Fits pool information regimes. Positive favors
 graph features. Intervals condition on retained questions; API panels have only
 two questions and receive point estimates only. Different API question composition
 precludes interpreting this table as a model leaderboard. $E_m$ abbreviates $E_m^{\mathrm{new}}$ in Eq.~\eqref{eq:mae-new}.}
 \label{tab:fresh}\centering\small
\begin{tabular}{llrll}
\toprule
Backend & Information & Questions & $\Delta\ell$ & $\Delta E_m$\\
\midrule
Qwen local 27.3B & global & 8 & $-0.0168\;[-0.0384, -0.0013]$ & $-0.0012\;[-0.0306, 0.0262]$\\
Qwen local 27.3B & local & 8 & $0.0039\;[-0.0206, 0.0265]$ & $-0.0101\;[-0.0343, 0.0121]$\\
Gemma local 8.0B & global & 8 & $-0.0121\;[-0.0278, 0.0012]$ & $-0.0048\;[-0.0097, 0.0003]$\\
Gemma local 8.0B & local & 8 & $-0.0057\;[-0.0244, 0.0098]$ & $-0.0005\;[-0.0056, 0.0035]$\\
Qwen Flash API & global & 2 & $0.0000$ & $0.0000$\\
Qwen Flash API & local & 2 & $0.0000$ & $0.0000$\\
DeepSeek Flash API & global & 2 & $-0.0005$ & $-0.0486$\\
DeepSeek Flash API & local & 2 & $0.0322$ & $-0.0649$\\
GLM API & global & 2 & $-0.0051$ & $-0.0046$\\
GLM API & local & 2 & $0.0291$ & $0.0054$\\
\bottomrule
\end{tabular}

\end{table*}

To test the shared-policy constraint, a separately frozen post-primary sensitivity
fits each information regime independently, preserving exactly the same balanced
panel and validation-only selection. For local-input Gemma, the graph advantage
becomes $\Delta\ell=\SeparateGemmaLocalLL$, while the collective comparison remains
unresolved, $\Delta E_m^{\mathrm{new}}=\SeparateGemmaLocalMAE$. Local-input Qwen has
$\Delta\ell=\SeparateQwenLocalLL$ and
$\Delta E_m^{\mathrm{new}}=\SeparateQwenLocalMAE$, both unresolved. The two-question
DeepSeek local panel changes direction collectively, favoring graph features under
separate calibration (Appendix~\ref{app:fresh}). Therefore the primary campaign's
unresolved graph comparison must not be read as evidence that local graph information is
unhelpful: pooled- and separate-regime calibration yield different estimates on
the same test panel. Nor does this sensitivity
establish that graph-aware policies always preserve collective dynamics.

Calibration sample size is fixed while test $N$ grows. Accordingly, an apparent
error plateau can include finite-calibration bias and regularization, not only
discarded interaction correlations. For single-class calibration, the smoothed
constant probability already differs from a deterministic consensus response.
These effects must be separated before interpreting an $N$ trend as closure error.

\section{Forecast horizon and calibration transfer}\label{sec:long}
\subsection{Calibration design and technical replay}
The three-round campaign leaves open whether the calibration-dependent results
persist over longer trajectories. A separately frozen follow-up uses the same two
pinned local installations and all four existing subjective questions, not a subset
chosen for large policy reversals. Every new episode has 12 updates. Training uses
$N=16$, replica 0; validation uses $N=32$, replica 0. Tests use $N=16,64$ and five
new graph/initial-state realizations, replicas 3--7. The design contains 32
calibration and 160 test episodes. Questions are shared across calibration and test;
these additional initializations do not constitute new-question generalization.

Two calibration strategies are fixed before collection. \emph{Short-frozen} uses
the exact 64 scalar, graph and one-/two-lag fits from the completed separate-regime
sensitivity. \emph{Long-refit} trains the same families on the new 12-round training
episodes, choosing regularization only on the new validation episodes. Both fit
each backend/question/regime separately. Recalibration changes sample count and
temporal coverage together. The short-frozen round covariate also extrapolates
beyond its original range. The comparison is not a causal decomposition of these
effects or an intervention on LLM memory.

The original run completes \LongOriginalComplete{} of 192 episodes. One Gemma
local-input test on question s2, $N=64$, replica 6 stops with a transport error and
no recorded response at round 5/node 25. The original matched analysis therefore
uses \LongOriginalPanels{} backend/question panels: four Qwen and three Gemma.
A dated technical-failure amendment then replays that entire episode from its
original initial condition, with the original model, decoding and sampling-seed
schedule. All 768 new responses pass the unchanged parser. An amended analysis,
called the replay overlay, includes the replacement trajectory and contains
\LongAmendedComplete{} complete cases and
\LongAmendedPanels{} panels, four questions per backend. The failed attempt and
original analysis remain intact; the replay is not an extra independent replica.
The figures below use this overlay. Appendix~\ref{app:long} compares the original
and amended panels and documents the reason for replay.

The prespecified evaluation horizons are $T=3,6,12$. We generalize
$E_m^{\mathrm{new}}$ by averaging through $T$ rather than three rounds, using 1,000
autonomous draws per initialization, common random numbers across policies, and
nested trajectory prefixes across horizons. Errors are averaged over replicas,
then equally over sizes and questions. With only four known questions, whole-question
bootstrap intervals are descriptive and fragile, not broad coverage guarantees.

\subsection{Local and collective effects of recalibration}
\begin{figure*}[t]
 \centering\includegraphics[width=\textwidth]{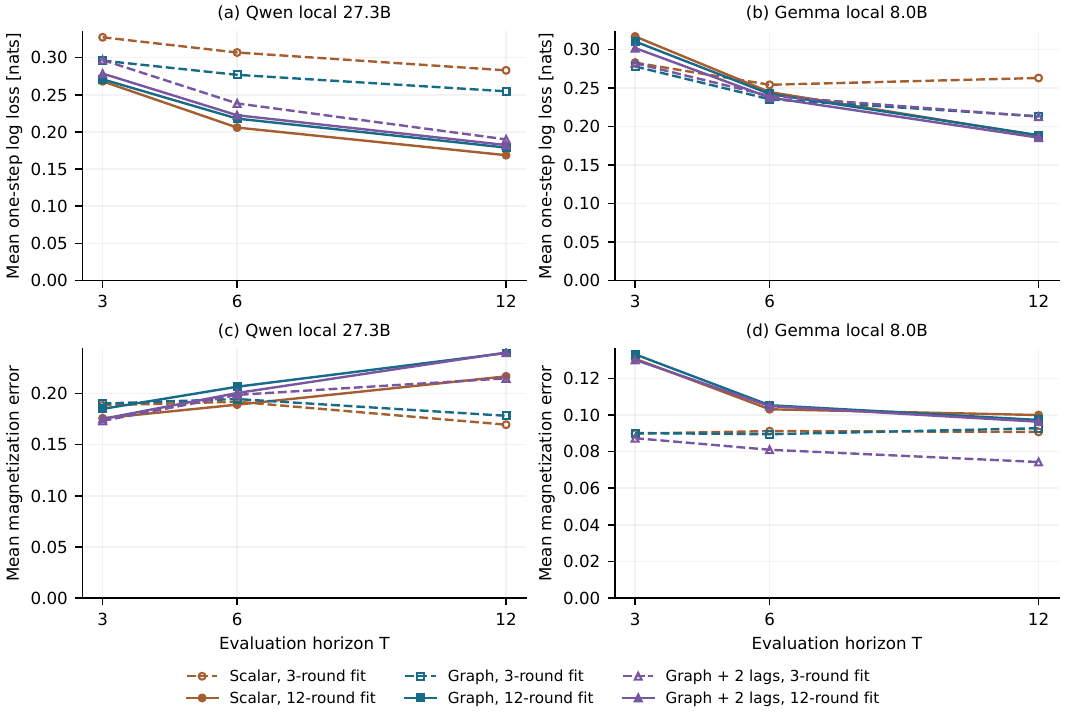}
 \caption{Completed 12-round follow-up, local-input regime. Top: observed-transition
 log loss; bottom: autonomous mean-magnetization error. Dashed lines use the
 original three-round fits; solid lines use independently generated 12-round
 calibration data with validation-only selection. Circle, square, and triangle
 markers denote scalar, graph, and graph-plus-history policies, respectively.
 Points average five test
 initializations, two sizes ($N=16,64$), and the same four known questions per
 backend. Horizons are nested, not independent experiments. Curves are descriptive
 point estimates, not uncertainty bands. This figure uses the technical-replay
 overlay; paired graph contrasts and replay sensitivity are given in
 Tables~\ref{tab:long} and~\ref{tab:retry}. $E_m$ abbreviates $E_m^{\mathrm{new}}$ in Eq.~\eqref{eq:mae-new}.}
 \label{fig:long}
\end{figure*}

Figure~\ref{fig:long} shows local-input scores. At $T=12$, Qwen scalar log loss
falls from $\LongQwenShortScalarLL$ to $\LongQwenRefitScalarLL$ under longer
calibration, a paired reduction of $\LongQwenScalarRefitLLGain$.
Its mean-trajectory error instead rises from $\LongQwenShortScalarMAE$ to
$\LongQwenRefitScalarMAE$; the paired reduction is
$\LongQwenScalarRefitMAEGain$. The latter interval includes zero: the local gain
is resolved within this panel, but a collective improvement is not. We do not
claim statistically resolved collective harm.

With short-frozen Qwen weights, adding two lags to the graph policy similarly
reduces log loss by $\LongQwenShortHistoryLLGain$, whereas its trajectory-error
reduction is $\LongQwenShortHistoryMAEGain$. Under long-refit weights, these
comparisons are both unresolved. For Gemma, scalar log loss changes from
$\LongGemmaShortScalarLL$ to $\LongGemmaRefitScalarLL$, while trajectory error
changes from $\LongGemmaShortScalarMAE$ to $\LongGemmaRefitScalarMAE$; its paired
calibration intervals both include zero. Table~\ref{tab:long} reports graph-versus-scalar
contrasts for both information regimes. There is no uniform graph advantage across
calibration choices. In particular, the resolved Qwen log-loss gain under longer
calibration does not imply a reduction in trajectory MAE on these four questions.
The twelve-update horizon does not test asymptotic stability or transfer to new questions.

\section{Question transfer under two calibration strategies}\label{sec:topics}
\subsection{Source-frozen and topic-calibrated evaluation}
The preceding long-horizon test reuses known questions. A new locally frozen
experiment instead retains all 20 preselected subjective statements, spanning such
topics as retail returns, sports officiating, data sharing, and public lighting.
They are new to this experiment, not necessarily to LLM pretraining, and are not
20 independently sampled semantic domains. The same two pinned local backends
generate six-round episodes in both information regimes. Per backend, question and
regime, training uses $N=16$, replica 0; validation uses $N=32$, replica 0; tests
use $N=16,64$, replicas 1 and 2. This gives 80 training, 80 validation and 320 test
episodes. Initial graphs and opinions are paired across backends and regimes.

We compare two fitting procedures. In the \emph{source-frozen} arm, each policy
is fitted on the four old subjective questions and then left unchanged. Fits
pool the original three-round training episodes separately for each backend and
input regime; validation episodes select regularization. All 16 source policies
are fixed before any new-topic LLM response. In the \emph{topic-calibrated} arm,
the same policy families are fitted separately for each new question, backend
and input regime. The designated training episode determines the weights,
and the validation episode selects regularization. Both arms predict exactly
the same withheld test episodes. Only the first arm tests a new question without
fitting to responses on that question.

The original grid has \TopicOriginalComplete{} complete cases and two failures.
One Gemma global-input test on u15, $N=64$, replica 2 lost a response to a transport
error. A separately frozen full-episode technical replay restores that case without
changing its designated calibration data. The other failure is a received Qwen
u09 local-input answer with an extra JSON field; it is retained, not repaired or
retried. Including the technical replacement gives \TopicComplete{} complete cases and
\TopicPanels{} balanced backend/question panels: all 20 for Gemma and 19 for Qwen.
Appendix~\ref{app:topics} preserves the original-versus-overlay comparison.

The primary contrast is graph versus scalar after topic-specific calibration, at
$T=6$ with local input. Scores average replicas, then sizes, then questions.
Two thousand whole-question bootstrap resamples produce descriptive intervals
conditional on the fixed questions and fitted policies. Source-frozen and history
comparisons, $T=3$ prefixes, and global-input results are secondary; their intervals
are not adjusted for multiple comparisons.

\subsection{Graph and history effects on new questions}
\begin{figure*}[t]
 \centering\includegraphics[width=\textwidth]{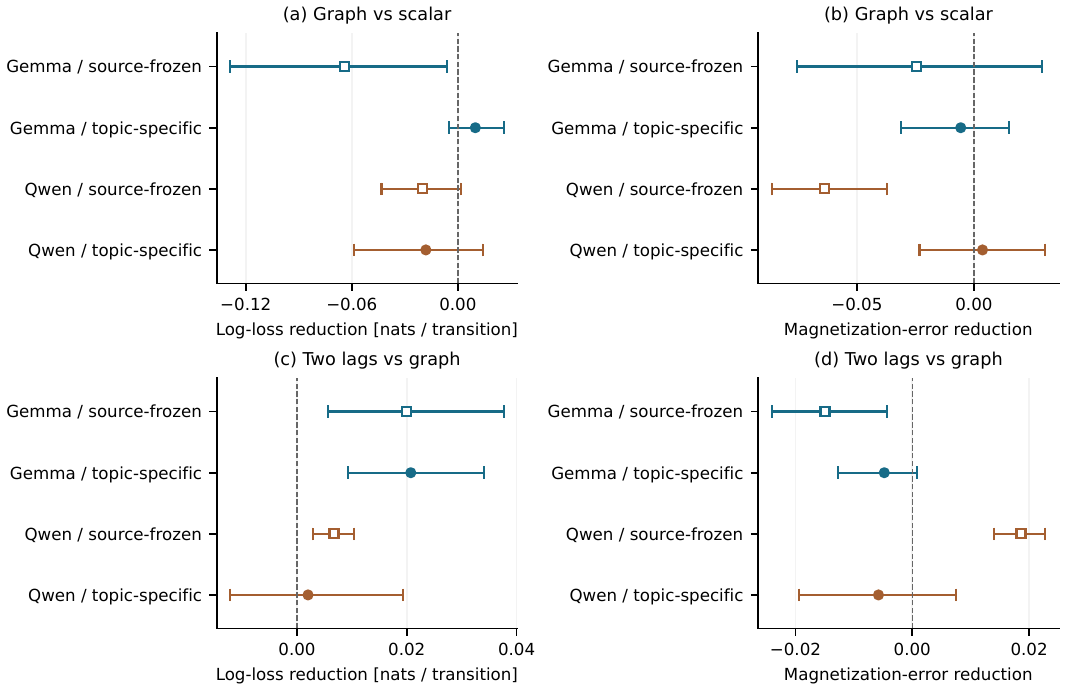}
 \caption{Completed new-question transfer test, local input at $T=6$. Top:
 graph-versus-scalar reductions; bottom: two-lag-versus-graph reductions. Left:
 observed-transition log loss; right: autonomous mean-magnetization error. Positive
 favors the alternative. Open squares are untouched source-frozen policies;
 filled circles use topic-specific training and validation, never test fitting.
 Points average two test realizations, two sizes and 20 Gemma or 19 Qwen questions;
 bars are descriptive 95\% whole-question bootstrap intervals. The primary
 topic-calibrated graph comparisons are unresolved. The secondary source-frozen
 Gemma history comparison improves the local score but worsens the collective
 score, while its Qwen counterpart improves both. This is the technical-replay
 overlay, not an additional independent experiment. The Gemma collective
contrast does not exclude zero under the broader multiplicity sensitivity
in Sec.~\ref{sec:multiplicity}; the error is Eq.~\eqref{eq:mae-new}.}
 \label{fig:topics}
\end{figure*}

Figure~\ref{fig:topics} contrasts local-input results. After topic-specific
calibration, neither backbone has a resolved graph-versus-scalar advantage in
log loss or trajectory MAE: all four primary intervals include zero.
Table~\ref{tab:topics} gives the paired numerical estimates and intervals. This
is neither an equivalence test nor evidence that graph information can never help.

One prespecified secondary comparison shows a more direct local--collective
reversal. In the source-frozen Gemma arm, adding two lags to graph features reduces
log loss by $\TopicGemmaFrozenHistoryLLGain$, but its trajectory-error reduction is
$\TopicGemmaFrozenHistoryMAEGain$: a resolved increase within these descriptive
intervals. The corresponding Qwen source-frozen comparison improves both metrics,
by $\TopicQwenFrozenHistoryLLGain$ and $\TopicQwenFrozenHistoryMAEGain$.
Under topic calibration, Gemma's two-lag log-loss gain remains positive but its
collective interval includes zero; both Qwen intervals include zero. We report
these contrasting outcomes rather than treating the Gemma reversal as universal.
The Gemma collective effect is sensitive to the separate post-hoc multiplicity
analysis in Sec.~\ref{sec:multiplicity}.

Calibration also changes absolute fidelity. For the local-input scalar policy,
Gemma trajectory MAE falls from $\TopicGemmaFrozenScalarMAE$ to
$\TopicGemmaAdaptScalarMAE$; Qwen's falls from $\TopicQwenFrozenScalarMAE$ to
$\TopicQwenAdaptScalarMAE$. Their paired reductions are
$\TopicGemmaScalarAdaptMAEGain$ and $\TopicQwenScalarAdaptMAEGain$.
This does not mean simply that more calibration data help: source fitting pools
192 training transitions, whereas a topic-specific fit uses only 96. Question
conditioning, temporal coverage and sample count change together. In particular,
source-frozen evaluation also extrapolates its round covariate beyond three
updates. Tables~\ref{tab:topics} and~\ref{tab:topicabsolute} give both regimes and
absolute errors, including unfavorable graph comparisons.

\section{Transfer to additional API backbones}\label{sec:api}
The local-model results prompted a separately frozen test on
\texttt{qwen3.8-max-0902} and \texttt{glm-5.2}, both served through Alibaba's
international endpoint. These are new observations from two API backbones,
not an inference about model scale from the earlier local installations.
The scientific grid was chosen solely from a small technical pilot's
availability, response compliance, latency and token usage. Within the fixed
budget, the predetermined selection rule retained both backbones, all twenty
statements and six rounds, but only $N=16$ for held-out evaluation.

For each backend, source policies were fitted on three-round trajectories
from the four old subjective questions and frozen before either backend's
first new-question request. Each new question then supplied one training
trajectory ($N=16$, replica 0), one validation trajectory ($N=32$, replica 0),
and one test trajectory ($N=16$, replica 1). Training observations determine
coefficients; validation log loss selects regularization. The test trajectory
is used only for evaluation. Untouched source policies and topic-specific
policies share that test trajectory. No collective error selects a fitted
model. The two calibration arms differ in topic, time coverage and training
count together, so their difference is not a pure sample-size effect.

The primary comparison, fixed before the new API responses, is the
topic-calibrated graph policy versus the scalar policy at $T=6$, separately
for each backbone. This comparison jointly adds degree and neighbor-state
features; the API grid has no separate degree-only control. We retain the same
two outcomes: observed-transition log
loss and the single-reference-path magnetization error of Eq.~\eqref{eq:mae-new}. We use
1,000 rollouts with common random numbers and 2,000 whole-question bootstrap
resamples. Source-frozen comparisons, history policies, the three-round
prefix and calibration contrasts are secondary. This is prospective transfer
to additional backbones after seeing local results, not a wholly unseen
hypothesis or an externally registered preregistration.

The original collection contains \APIOriginalComplete{} complete episodes
out of 120. One GLM validation episode stopped after two unanswered timeouts.
A separately frozen full technical replay, with unchanged model, prompts,
graph, initial state and decoding, supplies the replacement validation
trajectory. Its 192 responses are complete without transport or schema
errors. The amended collection contains \APIComplete{} complete episodes
and \APIPanels{} balanced backend--question panels. The failed original and
the original complete-panel analysis remain intact; the replay adds no test
replicate. Appendix~\ref{app:api} gives original-versus-amended sensitivity.

\begin{figure*}[t]
 \centering\includegraphics[width=\textwidth]{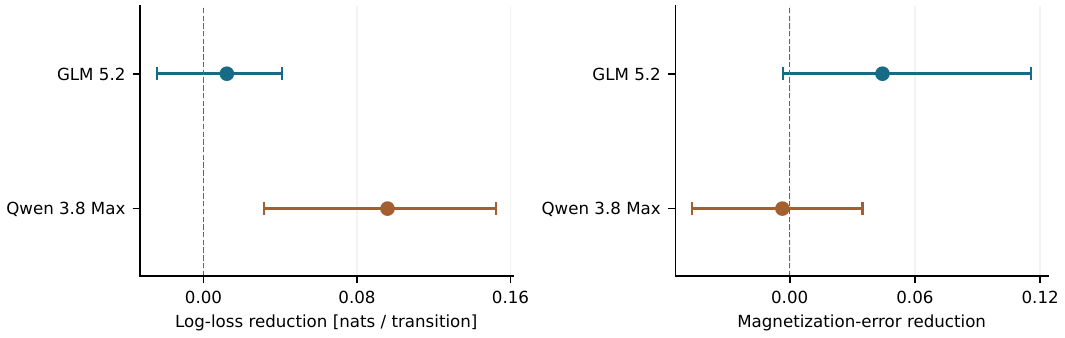}
 \caption{The same prespecified API contrasts as Table~\ref{tab:api-primary}.
 The dashed line marks zero improvement. Both local and collective scores
 are evaluated on the same held-out trajectories; a gain in one coordinate
 does not establish a gain in the other.}
 \label{fig:api}
\end{figure*}

Table~\ref{tab:api-primary} and Fig.~\ref{fig:api} report the fixed primary
comparison. Positive reductions favor the graph policy.

For GLM, the intervals do not clearly favor either policy for individual
predictions or group forecasts. For Qwen Max, graph information improves
individual prediction, but the group-forecast difference remains uncertain.
An interval containing zero does not establish equal performance or collective
harm. These statements use the reported nominal intervals; we examine the
effect of multiple-comparison adjustment below. The secondary
source-frozen Qwen graph comparison improves both scores: the log-loss
reduction is $0.1076\;[0.0696,0.1454]$ and the collective-error reduction is
$0.0759\;[0.0034,0.1458]$. Thus even within this backbone the conclusion depends
on which calibration arm is being evaluated. These are separate within-arm
contrasts, not a formal test of a difference between their effects. The
source-frozen collective interval includes zero under the broader post-hoc
multiplicity sensitivity in Sec.~\ref{sec:multiplicity}.
All six-round within-arm policy contrasts, including unresolved and adverse
estimates, are retained in Appendix~\ref{app:api}. Intermediate-horizon and
calibration-effect comparisons are retained in the machine-readable archive.
These intervals are descriptive and
conditional on the fixed questions, not multiplicity-adjusted guarantees.
One test trajectory per question also limits separation of surrogate bias
from stochastic variation in the observed population. In particular, this
fixed-$N$ follow-up cannot establish asymptotic scaling or a phase boundary.

A post-hoc descriptive audit finds single-class responses in 5 of 20 GLM
training episodes and 4 of 20 GLM test episodes, excluding imposed initial
spins from the counts. No Qwen training, validation or test episode is
single-class. No question is removed on this basis. This audit describes
response informativeness without changing the frozen analysis or treating
correlated node/time updates as independent samples.

\begin{table*}[t]
 \centering\small

\begin{tabular}{llrr}
\toprule
Model & $q$ & $\Delta\ell$ & $\Delta E_m$\\
\midrule
GLM 5.2 & 20 & $0.0122\;[-0.0242,0.0408]$ & $0.0444\;[-0.0030,0.1158]$\\
Qwen 3.8 Max & 20 & $0.0959\;[0.0313,0.1524]$ & $-0.0035\;[-0.0466,0.0349]$\\
\bottomrule
\end{tabular}

 \caption{Prespecified API-model comparison: topic-calibrated graph versus
 scalar policy, local input, $T=6$, held-out $N=16$. Entries are baseline-minus-
 alternative error reductions with descriptive 95\% whole-question bootstrap
 intervals; $q$ counts complete question panels after the validation replay.
 Local log loss is measured in nats per node transition. $E_m$ abbreviates $E_m^{\mathrm{new}}$ in Eq.~\eqref{eq:mae-new}.}
 \label{tab:api-primary}
\end{table*}

\section{Post-hoc sensitivity to the collective score}
\label{sec:proper-scores}
The single-path score in Eq.~\eqref{eq:mae-new} evaluates a predictive mean with
absolute loss. The distinction matters even with infinitely many surrogate draws:
if terminal magnetization is $+1$ with probability $0.9$ and $-1$ otherwise,
the true predictive mean $0.8$ has expected absolute error $0.36$, while a
misspecified deterministic forecast of $+1$ scores $0.20$. Squared loss instead
gives $0.36$ and $0.40$, respectively. This example diagnoses a scoring limitation,
not the cause of any particular empirical reversal~\cite{pointforecast}.

We therefore add a separate sensitivity analysis, specified after reviewing the
original results but before computing the new scores. It retains every saved
test policy, episode and horizon from the finite-grid, separate-regime,
long-horizon, local-topic and API-topic analyses, including original and
technically amended versions. The public study's reference-ensemble score and
terminal Wasserstein analysis are distinct and are not replaced. The fitted
weights, validation choices, exclusions, initialization, random seeds and
$R=1000$ draws remain unchanged. Only cheap autonomous rollouts are regenerated;
their original means, Monte Carlo standard errors and MAE are verified before
new scores are computed. No policy is fitted or selected using the new scores.

For episode $e$, define the ensemble mean
$\overline m_e(t)=R^{-1}\sum_r\widehat m_{er}(t)$ and sample variance
$s_e^2(t)=(R-1)^{-1}\sum_r[\widehat m_{er}(t)-\overline m_e(t)]^2$.
The fair mean squared error is
\begin{equation}
 E_{2,\mathrm f}(e)=\frac{1}{T}\sum_{t=1}^{T}
 \left\{[m_e(t)-\overline m_e(t)]^2-\frac{s_e^2(t)}{R}\right\}.
 \label{eq:fair-mse}
\end{equation}
For independent surrogate draws, subtracting $s_e^2/R$ removes the variance
of the Monte Carlo mean. Conditional on the observed path, this is unbiased for
the squared error of the surrogate's exact mean; averaging over reference paths
still includes irreducible reference variance. Negative finite-ensemble estimates
are retained. Uncorrected MSE is also saved.

To evaluate each round's predictive distribution, we use the fair ensemble
continuous ranked probability score (CRPS)~\cite{scoring,fairensemble}:
\begin{align}
 C_{\mathrm f}(e)&=\frac{1}{T}\sum_{t=1}^{T}\left\{
 \frac{1}{R}\sum_{r=1}^R|\widehat m_{er}(t)-m_e(t)|\right.\nonumber\\
 &\left.\quad-\frac{1}{R(R-1)}\sum_{r<s}
 |\widehat m_{er}(t)-\widehat m_{es}(t)|\right\}.
 \label{eq:fair-crps}
\end{align}
Here ``fair'' means unbiased over independent, identically distributed ensemble sampling for the population score
$\mathbb E|X-y|-\tfrac12\mathbb E|X-X'|$, with $X,X'$ independent surrogate
magnetizations and $y$ the observation. Its expected value is minimized by the
true marginal predictive law. Averaging marginal scores over time does not test
the full joint trajectory law or remove uncertainty from having few reference
paths.

Scores retain the original averaging order: replicas within size, sizes within
question, and equally weighted questions. Paired whole-question bootstrap
intervals use 2,000 resamples, with no interval below four questions. A further
post-hoc check uses 100,000 resamples and Bonferroni-style percentile limits for
all local-input, maximal-horizon, pooled-size policy contrasts and the two new
scores within each study/version. For the long, local-topic and API-topic studies,
this is 24 contrasts per study. Calibration effects remain descriptive, outside
that family. These intervals are conditional sensitivity diagnostics, not a
new confirmatory test or guaranteed family-wise coverage.

\begin{table*}[t]
\centering\footnotesize
\setlength{\tabcolsep}{3pt}
\begin{tabular}{lllrrr}
\toprule
Study/backend ($n_q$) & Arm & Baseline $\to$ alternative & $\Delta E_m$ & $\Delta E_{2,\mathrm f}$ [95\% interval] & $\Delta C_{\mathrm f}$ [95\% interval] \\
\midrule
Topics: Gemma (20) & TC & scalar $\to$ graph & $-0.0056$ & $-0.0084\;[-0.0358,+0.0116]$ & $-0.0043\;[-0.0283,+0.0158]$ \\
Topics: Qwen (19) & TC & scalar $\to$ graph & $+0.0037$ & $+0.0041\;[-0.0114,+0.0226]$ & $+0.0072\;[-0.0145,+0.0324]$ \\
API: GLM (20) & TC & scalar $\to$ graph & $+0.0444$ & $+0.0360\;[-0.0021,+0.1002]$ & $+0.0409\;[-0.0021,+0.1125]$ \\
API: Qwen Max (20) & TC & scalar $\to$ graph & $-0.0035$ & $-0.0166\;[-0.0584,+0.0150]$ & $-0.0016\;[-0.0337,+0.0294]$ \\
Topics: Gemma (20) & SF & graph $\to$ H2 & $-0.0149$ & $-0.0300\;[-0.0439,-0.0171]^{\dagger}$ & $-0.0221\;[-0.0312,-0.0117]^{\dagger}$ \\
Topics: Qwen (19) & SF & graph $\to$ H2 & $+0.0186$ & $+0.0226\;[+0.0147,+0.0318]^{\dagger}$ & $+0.0188\;[+0.0141,+0.0234]^{\dagger}$ \\
API: GLM (20) & SF & scalar $\to$ graph & $+0.0101$ & $-0.0250\;[-0.0905,+0.0409]$ & $-0.0091\;[-0.0633,+0.0408]$ \\
API: Qwen Max (20) & SF & scalar $\to$ graph & $+0.0759$ & $+0.1037\;[+0.0100,+0.1877]$ & $+0.0420\;[-0.0240,+0.1059]$ \\
Long: Qwen (4) & short & graph $\to$ H2 & $-0.0367$ & $-0.0276\;[-0.0539,+0.0004]$ & $-0.0338\;[-0.0694,+0.0133]$ \\
Long: Qwen (4) & refit & short $\to$ refit & $-0.0472$ & $-0.0317\;[-0.0604,+0.0041]$ & $-0.0366\;[-0.0810,+0.0261]$ \\
\bottomrule
\end{tabular}

\caption{Post-hoc collective-score sensitivity on amended balanced panels.
All ten displayed comparisons were fixed before computing these scores.
Input is local, sizes are pooled, and horizons are six rounds for topics/API
and twelve for the long study. Differences are baseline minus alternative;
positive values favor the alternative. Here $n_q$ counts questions, H2 denotes
two retained lags, SF source-frozen, TC topic-calibrated, and short the
short-frozen long-study arm. The last row
compares short-frozen with long-refitted scalar policies. Original MAE is
retained; brackets are paired 95\% descriptive intervals for fair MSE and CRPS.
A dagger marks intervals also excluding zero in the corresponding 24-contrast
post-hoc sensitivity; that adjustment does not apply to the last row.
All policies, horizons, sizes, original/amended versions and uncorrected MSE
are retained in the machine-readable analysis.}
\label{tab:collective-scores}
\end{table*}

After fitting separate policies for each question, none of the four
graph-versus-global comparisons shows a clear collective improvement under
either fair MSE or CRPS (Table~\ref{tab:collective-scores}). The earlier
history comparisons give a different result when policies fitted on the old
questions are left unchanged: adding two past rounds worsens Gemma forecasts
and improves local Qwen forecasts. Both directions persist after the
24-comparison adjustment. Gemma's fair-MSE and CRPS reductions are
$-0.0300\;[-0.0511,-0.0111]$ and $-0.0221\;[-0.0353,-0.0052]$;
local Qwen's are $+0.0226\;[+0.0114,+0.0373]$ and
$+0.0188\;[+0.0111,+0.0258]$. Thus Gemma's earlier forecast penalty is
not limited to mean absolute error. These additional scores were chosen
after inspecting the same questions and fitted policies; they are not an
independent replication.

For Qwen Max policies fitted only on the old questions, the graph improvement
depends on the score: the nominal fair-MSE interval is positive, whereas the
CRPS interval includes zero. Both adjusted intervals include zero. In the
long-horizon local Qwen study, the history and scalar-refitting comparisons
retain negative reductions with intervals crossing zero under both scores.
For the comparisons displayed here, restoring the technically failed local-topic
and API episodes does not change the directions or whether intervals include
zero. The complete numerical output also retains unfavorable graph comparisons:
for local Qwen policies fitted only on the old questions, graph features worsen
both collective scores relative to the global policy. Neither score supports
a consistent benefit from adding graph information or past opinions.

These results concern the earlier panels. The separately frozen new-statement
test in Sec.~\ref{sec:prospective} evaluates transfer of the local history and
adaptation findings without pooling new and old questions.

\subsection{Surrogate and message-content interventions}
\label{sec:mechanism-controls}
To investigate the Gemma--Qwen contrast, we subsequently fixed diagnostic
interventions on the common 19-topic panel, evaluating updates 4--6 with matched
random numbers. History is added to the numerical surrogate, not the LLM prompt.
Conditioning on the observed first three updates reverses Gemma's mean
graph--history ranking on the subsequent updates. Decomposing the original
simulator's signed forecast error shows that its generated past opinions
amplify the deviation toward agreement. Adjusting only the graph policy's
intercept using those three observed rounds sharply reduces Gemma's error,
but does not establish an advantage over a simple two-state model fitted
to the same rounds. These forecasts use extra observations; they do not
correct the original six-round forecasts from the initial state.

A separate intervention leaves the four original fits unchanged but caps their
raw round coordinate at its training maximum, $t=2$. For Qwen history-2 this
reduces late fair MSE from $\InvQwenClockBefore$ to $\InvQwenClockAfter$;
its history advantage becomes smaller but remains resolved in the declared
post-hoc sensitivity. Gemma does not improve under the same cap. Finally,
608 new local-model responses compare identical spin/graph contexts with native
versus stance-only messages. Agreement shifts differ between responders, but
Gemma's average effect is concentrated in Qwen-origin contexts, not its own.
On this previously inspected panel, response transfer, autonomous feedback and discarded text are distinct
diagnostic issues; the text test does not identify their contributions to the
original Gemma penalty. Appendix~\ref{app:history-investigation} gives the
interventions, negative controls and within-stage uncertainty families.

\section{Prospective replication on 24 additional statements}
\label{sec:prospective}
\subsection{Fixed predictors, adaptation rules and forecasting targets}
After the mechanism investigation, we fixed a new test before collecting its
first LLM response. It uses 24 additional subjective statements, the same pinned
local Gemma and Qwen installations, signed degree-four graphs, $N=16,64$, two
replicas and six synchronous updates: 192 episodes. All episodes and 46,080
node responses completed without a technical failure, retry or excluded question.
The protocol and executable plan were timestamped and hash-frozen on September
14, 2026, before collection; this is not a public external preregistration.
The statements have no normalized exact overlap with the previously recorded
statements, but are hand-selected, not random samples of domains. They are not
assumed to be unfamiliar from pretraining.

The original short-source graph and history-2 policies remain unchanged,
including coefficients, scaling and selected regularization. Autonomous
forecasts observe only state 0. Conditional forecasts observe states 0--3 and
predict updates 4--6, with the same original initial-spin features. Both are
scored only on updates 4--6; an observed prefix is never scored as a prediction.
The cap changes only the surrogate's raw time feature to $\min(t,2)$.
Conditional offset policies fit only the intercept correction of
Eq.~\eqref{eq:prefix-offset} to the observed prefix. Neither operation adds
history to the LLM prompt.

Two simple controls are fixed in advance. Persistence retains the last
available state. A two-state policy estimates the two probabilities of next
spin $+1$ conditional on current spin $\pm1$, with $1/2$ success and failure
pseudocounts. Its autonomous version uses only old source-training transitions;
its conditional version uses only the same episode's observed updates 1--3.
The conditional comparison thus evaluates specified adaptive procedures, not
zero-adaptation transfer. No future observation selects a policy or parameter.
Appendix~\ref{app:prospective} gives all forecast variants, collection rules,
absolute scores and exact statements.

The primary targets are fair MSE and fair marginal CRPS
(Eqs.~\eqref{eq:fair-mse} and~\eqref{eq:fair-crps}), using 1,000 surrogate draws
and matched absolute-round random numbers. Scores average rounds, then replicas
within size, sizes within question, and questions equally. All comparisons use
the same 24-question panel. The 28 predeclared contrasts cross two backends, two
scores and seven comparisons (Table~\ref{tab:prospective-primary}). The fixed
question bootstrap uses 100,000 resamples and percentile tails $0.025/28$ and
$1-0.025/28$~\cite{dunn,efron}. These are approximate adjusted intervals
conditional on the fixed panel and policies, not finite-sample coverage or
domain-generalization guarantees. An interval including zero is unresolved,
not evidence of equivalence. Nominal intervals and all secondary outputs remain
available, but do not replace this primary family.

\begin{table*}[t]
\centering\small
\setlength{\tabcolsep}{5pt}
\begin{tabular}{llrr}
\toprule
Backend & Contrast & $\Delta E_{2,\mathrm f}$ [adjusted interval] & $\Delta C_{\mathrm f}$ [adjusted interval] \\
\midrule
Gemma & AH & $-0.00117\;[-0.01496,+0.00643]$ & $+0.00783\;[-0.01580,+0.02864]$ \\
Gemma & AC & $-0.00053\;[-0.00195,+0.00024]$ & $-0.00048\;[-0.00308,+0.00188]$ \\
Gemma & AB & $+0.00115\;[-0.02309,+0.01428]$ & $+0.02042\;[-0.01977,+0.05442]$ \\
Gemma & CH & $+0.00224\;[-0.00318,+0.00649]$ & $+0.01264\;[-0.00702,+0.03080]$ \\
Gemma & CO & $+0.01006\;[-0.00859,+0.04740]$ & $+0.02939\;[-0.00298,+0.08119]$ \\
Gemma & CB & $+0.00756\;[+0.00001,+0.01789]$ & $+0.01441\;[-0.00134,+0.03320]$ \\
Gemma & GC & $+0.00341\;[-0.00045,+0.01628]$ & $+0.00481\;[-0.00073,+0.01753]$ \\
\midrule
Qwen & AH & $+0.01122\;[-0.00813,+0.03226]$ & $+0.00989\;[-0.00732,+0.02731]$ \\
Qwen & AC & $+0.02117\;[-0.07317,+0.12610]$ & $+0.01927\;[-0.05697,+0.09915]$ \\
Qwen & AB & $+0.15498\;[-0.04831,+0.33494]$ & $+0.17652\;[-0.00084,+0.34138]$ \\
Qwen & CH & $+0.02000\;[+0.00568,+0.03776]$ & $+0.02059\;[+0.00760,+0.03677]$ \\
Qwen & CO & $+0.02475\;[+0.00177,+0.06074]$ & $+0.02042\;[-0.00217,+0.05641]$ \\
Qwen & CB & $-0.13081\;[-0.19991,-0.06440]$ & $-0.20286\;[-0.29516,-0.11194]$ \\
Qwen & GC & $+0.00878\;[+0.00124,+0.01708]$ & $+0.01070\;[+0.00258,+0.01938]$ \\
\bottomrule
\end{tabular}

\caption{All 28 primary contrasts in the precollection-frozen 24-statement test,
updates 4--6, with equal question and population-size weights. For either error
score $S$, AH is autonomous $S(\mathrm{graph})-S(\mathrm{history2})$;
AC is autonomous $S(\mathrm{history2})-S(\mathrm{cap\ history2})$;
AB is autonomous $S(\mathrm{two\ state})-S(\mathrm{cap\ history2})$;
CH is conditional $S(\mathrm{graph})-S(\mathrm{history2})$;
CO is conditional $S(\mathrm{graph})-S(\mathrm{offset\ graph})$;
CB is conditional $S(\mathrm{two\ state})-S(\mathrm{offset\ graph})$;
GC is CH minus AH. Conditional forecasts observe states 0--3. Positive AH/CH
favors history; positive AC/CO favors the intervention; negative AB/CB favors
the simple baseline. All brackets use the same 28-contrast adjustment. Signs
relative to zero are evaluated before rounding.}
\label{tab:prospective-primary}
\end{table*}

\subsection{Autonomous and conditional transfer}
The earlier autonomous Gemma history penalty is not confirmed: adjusted
intervals include zero for both scores, and its CRPS point estimate changes
direction. Qwen's autonomous history advantage and the autonomous clock-cap
gain are also unresolved on this panel. These outcomes limit the generality
of the previous observations; they neither erase those observations nor show
that the underlying effects are zero. We do not pool panels, change the time
window or promote secondary comparisons to recover the earlier finding.

Conditional Qwen forecasts yield a different result. With three observed updates, history-2 reduces
mean fair MSE by $\ProsQwenHistoryMSEPercent\%$ and fair CRPS by
$\ProsQwenHistoryCRPSPercent\%$ relative to graph, with both adjusted intervals
positive. In comparison GC, the benefit of adding past opinions is larger after observing the
first three updates than when forecasting from the initial state. The adjusted
interval for this change is positive for both scores; both forecasts are scored
on updates 4--6. This changes forecasting information,
not the LLM's intrinsic memory. The graph offset gain is resolved
for Qwen fair MSE, but not CRPS.

Improvement within the graph-policy family does not establish superiority to
a simple forecast. Conditional Qwen two-state scores are
$\ProsQwenSimpleMSE$ and $\ProsQwenSimpleCRPS$, versus
$\ProsQwenOffsetMSE$ and $\ProsQwenOffsetCRPS$ for offset graph. The latter
errors are approximately $\ProsQwenBaselineMSERatio$ and
$\ProsQwenBaselineCRPSRatio$ times as large, respectively; both predeclared
baseline comparisons favor two-state after adjustment. This is evidence
against an advantage of this particular graph-retaining policy for these
targets and horizon, not a theorem that graph information is unnecessary.
For Gemma, offset graph's comparison with two-state has a positive fair-MSE
interval with lower bound only $\ProsGemmaBaselineMSELower$; its CRPS interval
includes zero. The near-zero MSE endpoint is also sensitive to Monte Carlo
estimation of the extreme bootstrap quantile
(Appendix~\ref{app:bootstrap-precision}). We retain the frozen interval;
this is not a robust advantage over two-state. All absolute forecast errors, including persistence
and both capped policies, are retained in Table~\ref{tab:prospective-absolute}.

On this panel, Qwen's conditional history gain has no resolved autonomous
counterpart, and its adaptive graph policy does not beat the two-state baseline.
The earlier opposite autonomous history effects for Gemma and Qwen and the
clock-cap gain are not confirmed. The experiment tests transfer of specified
predictors and adaptation procedures, not an asymptotic error exponent or the
minimal sufficient state of the underlying LLM society.

\section{Discussion}
\label{sec:synthesis}
\subsection{Local prediction and collective transfer}
Our simulations still track every agent. We test whether replacing LLM
responses with small statistical models preserves the selected group forecasts;
we do not replace the population with one variable. Public-data graph
features reduce one-step log loss in all 16 point estimates and reduce pooled
mean-trajectory MAE on held-out questions (Table~\ref{tab:pooled}). Under joint
question/graph shift, the primary MAE-reduction interval includes zero; after matched
semantic augmentation, the corresponding interval excludes zero on the positive side
(Table~\ref{tab:semantic}). Local rankings therefore
inform, but do not replace, direct collective evaluation.

We also fitted separate policies for each new question. Across two local and
two API backends, adding neighbor counts and opinions did not yield a
statistically clear improvement in group forecasts under fair MSE or CRPS
(Table~\ref{tab:collective-scores}). In earlier comparisons using unchanged
policies fitted on old questions, adding past opinions worsened Gemma forecasts
but improved local Qwen forecasts. Qwen Max's apparent graph benefit depends
on the score and multiple-comparison adjustment. These results do not establish
that history generally helps or harms, or that model size explains the
differences. Tables~\ref{tab:topics}, \ref{tab:api-primary}, and~\ref{tab:api-secondary}
retain the original estimates and intervals.

The prospective test does not confirm the autonomous history contrast or
clock-cap gains (Table~\ref{tab:prospective-primary}). For Qwen, adding past
opinions improves forecasts after three observed updates. A separate comparison
favors a two-state model fitted to those updates over the graph model with an
adjusted intercept, on both scores. These comparisons concern prediction of
updates 4--6 after observing updates 1--3, not a general explanation
of intrinsic Gemma--Qwen memory differences.

\subsection{Population size and scope of generalization}
The size comparisons use finite populations and specified calibration budgets.
Changing population size does not provide additional source observations for
an unchanged fitted policy. These experiments therefore do not isolate
finite-size sampling error from response-fitting error, or determine a
thermodynamic limit. Likewise, an increase in error over the tested grid for
a bounded observable is not divergence as $N\to\infty$.

Table~\ref{tab:studies} bounds the empirical claim: the finite grid is short and
includes single-class calibration tasks; longer horizons use four known
questions; topic panels are hand-selected, not randomly sampled domains, and
may be familiar from pretraining. Source-frozen and topic-specific arms test
different forms of transfer. The API study has one test size and one realization
per question; the 24-statement replication adds two sizes, not a large-size limit.
None identifies $N^{-1/2}$ convergence or an asymptotic error floor. Conditional
prefix forecasts also cannot substitute for autonomous simulation.

\subsection{Reference uncertainty and calibration scope}
The public analysis has 18 held-out questions and only two to four accepted
reference replicas per configuration. More surrogate draws reduce simulator
Monte Carlo error, not uncertainty in the LLM reference. In the public-data analysis, whole-question bootstrap
intervals condition on fitted models, fixed prompts, retained cases and the split;
they are descriptive and not multiplicity-adjusted coverage guarantees. An
interval crossing zero is unresolved, not evidence of equivalence. The
source-frozen and topic-calibrated arms share test trajectories, but their
different sample counts, topics and temporal coverage prevent a pure
sample-size interpretation of recalibration effects.

\subsection{Post-hoc multiplicity sensitivity}\label{sec:multiplicity}
A separate audit of the original log-loss and MAE outcomes resampled whole
questions 100,000 times and used
Bonferroni-style percentile limits~\cite{dunn,efron} at $0.025/24$ and $1-0.025/24$.
The family comprises 24 local-input, six-round contrasts within each
study: two backbones, two calibration arms, three policy comparisons
and two outcomes, pooling the tested sizes. It was selected post hoc
for sensitivity analysis, not as a uniquely prescribed testing family.
The source-frozen collective-error differences have intervals
$[-0.0282, +0.0027]$ for Gemma history-2 versus graph and
$[-0.0402, +0.1865]$ for Qwen Max graph versus scalar; positive differences
denote error reduction. Both include zero. With a narrower family
of four primary outcomes (two backbones, two outcomes, topic-calibrated
graph versus scalar), Qwen Max retains a resolved local gain but an
unresolved collective gain; its local interval also includes zero
under the broader 24-comparison family. These are approximate,
post-hoc sensitivity intervals conditional on the fixed questions
and fitted models, not guaranteed family-wise population coverage
and not a replacement for the frozen estimates and intervals above. The
additional squared-error and CRPS analysis in Sec.~\ref{sec:proper-scores} uses
a separate, explicitly declared sensitivity family; the MAE conclusion must not
be generalized to every collective score.

\subsection{Selection effects and parameter identifiability}
The integrity filter and schema failures can select question panels nonrandomly.
Matched accepted cases permit within-panel policy comparisons but do not identify
missing valid outcomes. Technical replays restore only named failed episodes
under separate amendments; original results and failed attempts remain visible.
They are not new test replicates. Different model panels must not be read as
a paired quality leaderboard. Public graph transfer changes signs, degrees and
questions together, and the new global/local prompting changes semantic content
as well as information routing. Neither is a causal topology-only intervention.

Our public-data estimators omit or differently encode information used by the
source authors. Their absolute errors are not evidence against the accuracy of
the authors' richer model. Likewise, conditional total correlation in
Eq.~\eqref{eq:path} is a possible unmeasured error source, not an experimentally
identified cause of the observed ranking differences. The lag-policy comparisons
measure forecasting effects, not one shared memory kernel across domains.

The three-update source calibration has a further identification limit.
With the initial padding in Eq.~\eqref{eq:history-features}, the second-lag
spin is $s_i(\max\{0,t-2\})=s_i(0)$ at all predictor times $t=0,1,2$.
Let $a_0$ and $a_2$ be the raw-logit coefficients of the initial spin and
second-lag spin, respectively. For any real $c$, the transformation
\begin{equation}
 (a_0,a_2)\longmapsto(a_0+c,a_2-c)
 \label{eq:lag-alias}
\end{equation}
leaves every training and validation likelihood unchanged. After the third
update the two inputs can differ, so equivalent short-source predictors can
produce different autonomous continuations. A regularizer selects an extension;
the short histories do not identify that extension. This specific ambiguity
need not explain the entire collective-score difference. Longer source histories
can separate these inputs without guaranteeing transfer. Thus the
Gemma--Qwen contrast is a comparison of fitted simulators, not a causal test of intrinsic LLM memory.
The subsequent controls in Appendix~\ref{app:history-investigation} isolate
specific sensitivities of those simulators on the earlier panel, including Qwen's extrapolated clock
and Gemma's lag feedback; the prospective results limit the transfer of those
diagnoses. Their new text responses establish a responder
difference on matched contexts, but not a unique explanation of the earlier
collective ranking or a pure architecture/size effect.

\subsection{Validation criteria and open questions}
For future applications, a validation protocol should specify an acceptable
error tolerance, the observable, reference ensemble and horizon,
and choose candidate information sets before fitting. It separates training, validation and
test questions or trajectories at the level appropriate to the transfer claim.
It evaluates both a proper observed-transition score and autonomous collective
outcomes, using scores consistent with the target mean or distribution, a common
test panel and explicit uncertainty units. Autonomous simulation and forecasting
after an observed prefix are separate tasks: report both the observations and
the permitted adaptation. Compare with prespecified persistence and simple
transition baselines using the same information budget; a within-family gain
does not establish practical value. A simpler
state should be preferred only if its held-out error and uncertainty satisfy
the stated tolerance; an information-topology label alone cannot establish this.

Stronger claims would require additional evidence: new questions drawn from a
prespecified sampling frame, independent reference replicas for each initialized
system to evaluate fluctuation distributions,
content-controlled multi-round interventions, and size sweeps that vary calibration budget
separately from $N$. Held-out longer histories are needed to test memory
truncation. The exploratory single-step text intervention and the fixed-statement
replication do not supply these stronger tests.
The broader economic and social case studies in Ref.~\cite{poor} motivate the
research question but are not empirical evidence for the comparisons reported here.

\section{Conclusion}
Graph information improves individual prediction in all public-data point
estimates and improves pooled group forecasts on held-out questions. That
group benefit does not hold consistently across the tested transfer settings.
The 24 additional statements do not confirm the earlier opposite effects of
past opinions on Gemma and Qwen forecasts from the initial state. Qwen does
benefit from past opinions when forecasting after three observed rounds.
For Qwen, a simple two-state model fitted to the first three observed rounds
also beats the graph model with an adjusted intercept on both collective scores.
This comparison does not cover every graph or history model.

Accurate individual prediction is therefore insufficient evidence of accurate
group forecasts.

Evaluation must specify the group quantities of interest,
the observations available to each predictor and any permitted refitting.
It must include simple baselines and uncertainty estimates. Our results
neither reduce the whole population to a few variables nor identify intrinsic
memory differences between LLM families. In particular, the short training
histories cannot uniquely determine all history coefficients.

\section*{Data, code, and relation to prior work}
This is a standalone empirical companion to the arXiv preprint
Poor Man's Agentic Modeling~\cite{poor}. That preprint motivates the
information-and-memory comparisons; its coarse-graining derivations and
fitted-response experiments are not reproduced here. This article reports
public-data reanalysis, separately collected LLM trajectories, exploratory
controls and the subsequent 24-statement test. These evidence sources are
distinguished throughout, and overlapping questions are not independent
replications.

The upstream trajectories and code are public~\cite{dataset,code}, with their
pinned revisions documented in Appendix~\ref{app:audit}. Local experiment archives
retain frozen protocols, input hashes, fitted coefficients, validation grids,
per-question outcomes, bootstrap summaries, raw response receipts, exclusions and
technical amendments. Original and replay-amended analyses remain separate.
OpenRouter's incomplete technical pilots do not enter the scientific comparisons.
The exact new-campaign prompts, persona roster and statements are included in
Appendices~\ref{app:protocol-inputs} and~\ref{app:prospective}. A separate source-only reproducibility
package is maintained in a private repository. It contains the numerical code,
frozen protocols, required raw observations and minimal failure evidence;
historical fitted models, scores, tables and PDFs are not input dependencies.
Separate deposits retain all 608 new one-step text responses and the frozen
plan and receipts for the 192 prospective episodes;
their analyses do not replace the earlier panels. Access to the private package
can be requested from the corresponding author; no public release is asserted.

Offline reanalysis reproduces estimates from archived responses, not fresh LLM
generation. Local aliases and artifact digests identify the installed backends
but do not establish independently verified official model identities or provide
the model weights. Exact re-generation requires access to those artifacts and
the recorded runtime and decoding settings; live API versions need not be
immutable. The package documents these separate reproducibility targets and
reviewer entry points in \texttt{REVIEWER\_ACCESS.md}.

AI tools, including OpenAI Codex, were used to assist with code development,
analysis, and manuscript preparation. The research software was checked using
automated tests and end-to-end numerical reproduction from archived inputs.
The author takes responsibility for the scientific content and conclusions.

Offline entry points reproduce the public and new-campaign fits and rollouts
(\texttt{reproduce.py}) and the frozen prospective analysis
(\texttt{prospective\_topics\_analysis.py}). The reporting pipeline
\texttt{prospective\_paper.py} generates numerical tables and figures from
these analyses; \texttt{standalone\_paper.py} assembles this article from its
separate manuscript source and hash-verified reporting artifacts, without new
inference or fitting. The analysis package also contains material used by the
earlier framework, which is not included in this article. Input hashes,
numerical regression checks and receipt audits link estimates to observations.
Recompiling a PDF alone does not reproduce the analysis. Tests check
implementation and provenance, not empirical validity. Formal checks supplied
with the broader project do not verify the whole article or its empirical
conclusions.

\onecolumngrid
\raggedbottom
\appendix

\section{Public-data inclusion criteria and numerical implementation}\label{app:audit}
The public-data analysis uses the fixed dataset and code snapshots cited in
Refs.~\cite{dataset,code}. Full revision identifiers, file checksums and campaign
fingerprints are recorded in the reproducibility package's
\texttt{APPENDIX\_PROVENANCE.md} and machine-readable manifests.

The integrity check finds 5,214 raw zero votes, 53 zero-valued recorded node-time
states, and 120 node-time disagreements between the recorded state and the raw
majority. These are different counting units: several problematic votes or states
can occur in one episode. The strict filter first rejects complete histories with
nonbinary states, then rejects any remaining episode with a majority disagreement.
Table~\ref{tab:audit} reports the resulting episode counts, not the counts of
individual faulty votes.

\begin{appendixtable}
 \caption{Complete episodes accepted by the integrity filter. Original counts are
 1,600 objective and 800 subjective episodes for each backbone.}
 \label{tab:audit}\centering
 \begin{tabular}{lrrr}\toprule
 Backbone & Objective & Subjective & Excluded\\\midrule
 GPT-4o-mini & 1600 & 800 & 0\\
 Qwen3.5-9B & 1597 & 800 & 3\\
 Gemma-3n-E4B-it & 1520 & 738 & 142\\
 Llama-3.1-8B & 1600 & 800 & 0\\\midrule
 Total & 6317 & 3138 & 145\\\bottomrule
 \end{tabular}
\end{appendixtable}

The question-split seed is 260906, primary rollout seed 260907, and question-bootstrap
seed 260908. Episode-specific rollout streams add a hash of backbone, question, graph,
and replica to the base seed. The Monte Carlo diagnostic uses seed 260909 and was
specified after the primary results were available. It is stored separately and does
not replace the original estimates.

The implementation uses double-precision logistic optimization with analytic
gradients and L-BFGS-B~\cite{byrd}. All selected primary fits converge. Automated
checks cover grouped data separation, signed-field orientation, node-permutation
equivariance, exclusion of future observations from predictors, train-only
standardization, autonomous feedback, metric aggregation and model restoration.
Software versions and numerical regression checks accompany the code; these
verify implementation and reproduction, not the empirical hypothesis.

\Needspace{12\baselineskip}
\section{Complete public-data scores}
Table~\ref{tab:cells} gives absolute scores for all 16 combinations of backbone,
question type and transfer setting, rather than only their pooled differences.
The three policies use the same accepted test episodes within each combination.
Graph-policy four-way balanced accuracy ranges from 0.717 to 0.922; these values
are not directly comparable with published headline accuracies based on different
splits, covariates or target processing. The table separates observed-transition
log loss, mean-trajectory error and terminal-distribution distance because a
policy need not improve all three.

\begin{appendixtable}
 \caption{Absolute primary-run test scores; lower is better. History is the
 validation-selected depth (two lags in all cells). O/S denote objective/subjective
 questions. These node-resolved policies share current spin, initial spin, and round.
 Results for interaction-free, static-degree, normalized-neighbor, one-lag, and
 persistence controls, including Brier and absolute-magnetization scores, are
 retained in the accompanying machine-readable results.}
 \label{tab:cells}\centering\small
\begin{tabular}{lrrrrrrrrr}
\toprule
 & \multicolumn{3}{c}{One-step log loss} & \multicolumn{3}{c}{Mean-trajectory MAE} & \multicolumn{3}{c}{Terminal $W_1$}\\
Backbone / mode & Global & Graph & History & Global & Graph & History & Global & Graph & History\\
\midrule
\multicolumn{10}{l}{\emph{Held-out questions}}\\
GPT / O & 0.567 & 0.215 & 0.208 & 0.370 & 0.336 & 0.327 & 0.593 & 0.495 & 0.470\\
GPT / S & 0.506 & 0.197 & 0.178 & 0.206 & 0.166 & 0.162 & 0.279 & 0.238 & 0.234\\
Qwen / O & 0.318 & 0.199 & 0.190 & 0.519 & 0.489 & 0.510 & 0.604 & 0.559 & 0.578\\
Qwen / S & 0.460 & 0.302 & 0.285 & 0.176 & 0.127 & 0.130 & 0.249 & 0.202 & 0.206\\
Gemma / O & 0.282 & 0.256 & 0.236 & 0.275 & 0.270 & 0.268 & 0.299 & 0.295 & 0.296\\
Gemma / S & 0.284 & 0.221 & 0.202 & 0.348 & 0.336 & 0.338 & 0.388 & 0.378 & 0.385\\
Llama / O & 0.279 & 0.178 & 0.174 & 0.223 & 0.212 & 0.214 & 0.291 & 0.254 & 0.254\\
Llama / S & 0.227 & 0.135 & 0.126 & 0.399 & 0.283 & 0.262 & 0.520 & 0.357 & 0.320\\
\midrule
\multicolumn{10}{l}{\emph{Held-out questions + unsigned lattices}}\\
GPT / O & 0.457 & 0.121 & 0.122 & 0.398 & 0.397 & 0.391 & 0.567 & 0.554 & 0.541\\
GPT / S & 0.483 & 0.189 & 0.170 & 0.211 & 0.212 & 0.186 & 0.267 & 0.348 & 0.306\\
Qwen / O & 0.258 & 0.146 & 0.141 & 0.595 & 0.490 & 0.519 & 0.675 & 0.538 & 0.567\\
Qwen / S & 0.331 & 0.211 & 0.192 & 0.246 & 0.152 & 0.120 & 0.304 & 0.225 & 0.186\\
Gemma / O & 0.225 & 0.200 & 0.177 & 0.304 & 0.303 & 0.296 & 0.380 & 0.356 & 0.357\\
Gemma / S & 0.210 & 0.140 & 0.137 & 0.450 & 0.439 & 0.425 & 0.512 & 0.553 & 0.525\\
Llama / O & 0.212 & 0.100 & 0.096 & 0.234 & 0.195 & 0.192 & 0.315 & 0.218 & 0.226\\
Llama / S & 0.183 & 0.145 & 0.127 & 0.421 & 0.531 & 0.488 & 0.542 & 0.672 & 0.625\\
\bottomrule
\end{tabular}

\end{appendixtable}

\Needspace{12\baselineskip}
\section{Public-data split and integrity sensitivity}\label{app:extensions}
Table~\ref{tab:robustness} reports all 14 pooled graph-versus-global comparisons
from the five additional split seeds and two alternative integrity rules, across
both transfer settings. Each pooled estimate averages backbones within question
before bootstrap resampling. The majority-reconstructed rows reproduce the primary
population and outcomes and must not be counted as an independent confirmation.
Machine-readable artifacts also retain common-reference evaluations, history
comparisons, all validation grids and the complete semantic PCA transforms.

\begin{appendixtable}
 \caption{Post-primary split and integrity sensitivity: graph-versus-global paired
 reductions with 95\% question-cluster intervals. Each setting has 18 test questions;
 different seeds reuse the underlying 60-question corpus. ``Recorded binary'' omits
 the majority-agreement requirement; ``Majority'' yields the same accepted histories
 as the primary strict rule on this snapshot.}
 \label{tab:robustness}\centering\small
\begin{tabular}{llrr}
\toprule
Policy / split seed & Transfer & $\Delta\ell$ & $\Delta E_m$\\
\midrule
Majority / 260906 & Questions + lattices & $0.1412\;[0.1216, 0.1635]$ & $0.0238\;[-0.0059, 0.0556]$\\
Majority / 260906 & Questions & $0.1518\;[0.1370, 0.1681]$ & $0.0315\;[0.0188, 0.0465]$\\
Recorded binary / 260906 & Questions + lattices & $0.1416\;[0.1219, 0.1641]$ & $0.0240\;[-0.0060, 0.0557]$\\
Recorded binary / 260906 & Questions & $0.1516\;[0.1369, 0.1679]$ & $0.0319\;[0.0192, 0.0469]$\\
Strict / 260910 & Questions + lattices & $0.1333\;[0.0905, 0.1711]$ & $0.0250\;[-0.0281, 0.0735]$\\
Strict / 260910 & Questions & $0.1580\;[0.1303, 0.1816]$ & $0.0078\;[-0.0037, 0.0191]$\\
Strict / 260911 & Questions + lattices & $0.1576\;[0.1246, 0.1880]$ & $-0.0034\;[-0.0671, 0.0516]$\\
Strict / 260911 & Questions & $0.1685\;[0.1453, 0.1910]$ & $0.0062\;[-0.0052, 0.0173]$\\
Strict / 260912 & Questions + lattices & $0.1388\;[0.1112, 0.1665]$ & $0.0185\;[-0.0195, 0.0604]$\\
Strict / 260912 & Questions & $0.1565\;[0.1383, 0.1746]$ & $0.0198\;[0.0077, 0.0321]$\\
Strict / 260913 & Questions + lattices & $0.1519\;[0.1230, 0.1824]$ & $0.0002\;[-0.0466, 0.0409]$\\
Strict / 260913 & Questions & $0.1666\;[0.1444, 0.1876]$ & $0.0043\;[-0.0076, 0.0156]$\\
Strict / 260914 & Questions + lattices & $0.1359\;[0.1065, 0.1651]$ & $0.0315\;[-0.0238, 0.0817]$\\
Strict / 260914 & Questions & $0.1611\;[0.1365, 0.1845]$ & $0.0233\;[0.0121, 0.0338]$\\
\bottomrule
\end{tabular}

\end{appendixtable}

\Needspace{12\baselineskip}
\section{Finite-grid models, response processing and calibration}\label{app:fresh}
\subsection{Model and generation settings}
The local Qwen installation reports the alias \texttt{qwen3.8:latest}, parent
\texttt{qwen3.8:27b-q4\_K\_M}, 27.3B parameters and Q4\_K\_M quantization.
The local Gemma installation reports \texttt{gemma4:9b}, 8.0B parameters and
Q4\_K\_M quantization. These names are installed metadata, not independently
verified official model identities. Exact model digests are pinned in the
archived protocol.
Both servers report Ollama 0.30.0. API identifiers are
\texttt{qwen3.5-flash-2026-02-23}, \texttt{deepseek-v4-flash}, and
\texttt{glm-5.2}; version immutability beyond these returned identifiers is not
established.

Generation uses temperature 0.7, an output cap of 128 tokens, JSON mode and disabled
thinking. Local requests use a 4,096-token context and a sample-identity-derived
seed. Each answer must have exactly \texttt{opinion} (integer $\pm1$) and a nonempty
\texttt{message}, at most 600 characters and 64 words. Messages beyond the requested
25 words are retained with a style warning under the frozen campaign rule.
No neutral vote is recoded as a binary spin. Technical pilot responses are
excluded from scientific fitting.

\subsection{Missing episodes and calibration comparison}
Complete/planned episode counts are 160/160 for each local backend, 40/48 for Qwen
Flash, 46/48 for DeepSeek and 24/24 for GLM. Failures in training or validation exclude
a backend/question from fitting rather than triggering replacement calibration.
All failures and successful early rounds remain in the original archive, although
incomplete episodes do not enter the primary metrics. The balanced-panel policy
removes both regimes and all sizes together for any affected test question.

Of 440 planned episode records, 430 are complete. Complete calibration permits
23 backend/question pairs; the primary balanced test panel contains 304 episodes.
Seven further complete test episodes belong to an incompletely observed question
panel and are not substituted into that balanced comparison.

Table~\ref{tab:regime} compares a common fitted response pooled across information
regimes with separately calibrated global- and local-input responses on exactly
the same balanced panel. This sensitivity was specified after the primary
results were inspected. In both versions, standardization uses training data
only and regularization is chosen on designated validation episodes. The table
reports within-backend policy differences, not a ranking of model providers.
Complete intervals and history-policy results accompany the numerical outputs.

\begin{appendixtable}
 \caption{Post-primary calibration sensitivity on the unchanged balanced test
 panel. Entries are graph-versus-scalar point reductions; positive favors graph
 features. Separate-regime fitting can change both the sign and magnitude of a
 difference. $\Delta\ell$ is the log-loss reduction and $\Delta E_m$ the reduction
 in Eq.~\eqref{eq:mae-new}.}
 \label{tab:regime}\centering\small
\begin{tabular}{llrrrr}
\toprule
 & & \multicolumn{2}{c}{Pooled-regime fit} & \multicolumn{2}{c}{Separate-regime fit}\\
Backend & Information & $\Delta\ell$ & $\Delta E_m$ & $\Delta\ell$ & $\Delta E_m$\\
\midrule
Qwen local 27.3B & global & -0.0168 & -0.0012 & -0.0119 & -0.0123\\
Qwen local 27.3B & local & 0.0039 & -0.0101 & 0.0160 & 0.0016\\
Gemma local 8.0B & global & -0.0121 & -0.0048 & -0.0066 & -0.0033\\
Gemma local 8.0B & local & -0.0057 & -0.0005 & 0.0169 & -0.0022\\
Qwen Flash API & global & 0.0000 & 0.0000 & 0.0000 & 0.0000\\
Qwen Flash API & local & 0.0000 & 0.0000 & 0.0000 & 0.0000\\
DeepSeek Flash API & global & -0.0005 & -0.0486 & -0.2029 & -0.0794\\
DeepSeek Flash API & local & 0.0322 & -0.0649 & 0.1858 & 0.0687\\
GLM API & global & -0.0051 & -0.0046 & 0.0000 & 0.0000\\
GLM API & local & 0.0291 & 0.0054 & 0.0293 & 0.0004\\
\bottomrule
\end{tabular}

\end{appendixtable}

\Needspace{12\baselineskip}
\section{Long-horizon results and missing-case sensitivity}\label{app:long}
The 12-round follow-up preserves the prompts, signed-ring construction, personas
and decoding settings of the three-round campaign. Its design was frozen before
the new responses but after inspection of the earlier results. All four existing
subjective statements are retained; excluding the often single-class objective
tasks limits the population to which the results apply. The 32 calibration and
160 test episodes are newly generated. Test replicas are new graph/initial-state
realizations, not repeated simulations of an identical initial condition.

Table~\ref{tab:long} compares short-frozen policies, fitted on three-round data,
with long-refit policies, fitted on designated 12-round training and validation
episodes. Neither arm fits to test outcomes. Longer calibration changes both
sample count and temporal coverage; it does not isolate either effect.

\subsection{Technical replay and panel restoration}
The original run completes 191 of 192 episodes. The missing Gemma/s2 local-input
test ($N=64$, replica 6) stops at round 5/node 25 with a transport error and no
response body. A dated amendment on 8 September 2026 permits a full replay of
this named failure. Received schema-invalid or scientifically inconvenient
answers do not qualify for replay.

The replay keeps the original model, prompt, decoding and per-node sampling
identities, but is stored as a separate attempt. Matching seeds does not ensure
bitwise-identical GPU output: the replay differs from the original completed
prefix. The replacement is therefore a complete trajectory, never a new suffix
spliced into the old path. All 768 replay responses pass the unchanged parser.
The failed attempt and original analysis remain preserved.

The restored episode is test data, so neither training nor validation changes.
It restores the s2 matched panel, expanding Gemma's comparison from three to four
questions; Qwen's four-question panel is unchanged. Table~\ref{tab:retry} compares
the results before and after including s2. Because the question set changes,
the difference does not isolate an effect of replaying one trajectory. The
replay adds neither an independent test realization nor a previously unseen question.

\begin{appendixtable}
 \caption{Twelve-round graph-versus-scalar paired reductions, averaged equally over
 $N=16,64$. Positive favors graph features. All entries use the technically amended
 four-question panel for each backend and five test initializations per size. Brackets
 are descriptive 95\% whole-question bootstrap intervals. The same questions occur
 in both calibration arms; no test outcome selects fits or regularization. $E_m$ abbreviates $E_m^{\mathrm{new}}$ in Eq.~\eqref{eq:mae-new}.}
 \label{tab:long}\centering\small
\begin{tabular}{lllrr}
\toprule
Backend & Input & Fit horizon & $\Delta\ell$ & $\Delta E_m(12)$\\
\midrule
Qwen & global & 3 rounds & $1.1655\;[-0.0309, 3.2847]$ & $0.0073\;[-0.0172, 0.0381]$\\
Qwen & global & 12 rounds & $-0.0290\;[-0.0545, -0.0036]$ & $0.0202\;[-0.0368, 0.1037]$\\
Qwen & local & 3 rounds & $0.0282\;[-0.0060, 0.0708]$ & $-0.0087\;[-0.0296, 0.0027]$\\
Qwen & local & 12 rounds & $-0.0102\;[-0.0456, 0.0203]$ & $-0.0228\;[-0.0628, 0.0171]$\\
Gemma & global & 3 rounds & $-0.0010\;[-0.0196, 0.0162]$ & $-0.0049\;[-0.0328, 0.0153]$\\
Gemma & global & 12 rounds & $0.0056\;[-0.0023, 0.0134]$ & $-0.0028\;[-0.0057, 0.0002]$\\
Gemma & local & 3 rounds & $0.0497\;[-0.0001, 0.1414]$ & $-0.0020\;[-0.0154, 0.0079]$\\
Gemma & local & 12 rounds & $-0.0007\;[-0.0176, 0.0117]$ & $0.0027\;[-0.0058, 0.0125]$\\
\bottomrule
\end{tabular}

\end{appendixtable}

\begin{appendixtable}
 \caption{Technical-replay sensitivity for Gemma at $T=12$: graph-versus-scalar
 point reductions before and after restoring the transport-failed case. Original
 estimates omit all s2 tests from the comparison; the amended analysis restores
 them. The question set changes from three to four, so the difference does not
 isolate the effect of replaying one trajectory. Original three-question intervals are
 suppressed; amended intervals appear in Table~\ref{tab:long}. Qwen is unchanged. $E_m$ abbreviates $E_m^{\mathrm{new}}$ in Eq.~\eqref{eq:mae-new}.}
 \label{tab:retry}\centering\small

\begin{tabular}{llrrrr}
\toprule
 & & \multicolumn{2}{c}{Original ($q=3$)} & \multicolumn{2}{c}{Replay ($q=4$)}\\
Input & Fit rounds & $\Delta\ell$ & $\Delta E_m$ & $\Delta\ell$ & $\Delta E_m$\\
\midrule
global & 3 & -0.0020 & -0.0067 & -0.0010 & -0.0049\\
global & 12 & 0.0044 & -0.0024 & 0.0056 & -0.0028\\
local & 3 & 0.0628 & -0.0036 & 0.0497 & -0.0020\\
local & 12 & 0.0078 & 0.0065 & -0.0007 & 0.0027\\
\bottomrule
\end{tabular}

\end{appendixtable}

\Needspace{12\baselineskip}
\section{New-question results and missing-case sensitivity}\label{app:topics}
\subsection{Calibration and evaluation}
The new-topic design was frozen on 7 September 2026 before any new-question
response, but after the earlier campaign had been inspected. Its claims are
prospective within this follow-up, not externally registered confirmation of an
unseen hypothesis. Exact normalized statements were checked against the public
dataset and old campaign; the archived wording and topic IDs specify the test
population. No question was dropped for an inconvenient scientific result.

The source-frozen arm uses the original s1--s4 training/validation episodes, not
new elicitation on these topics. Its four families are fit independently for each
backend/regime; no current-topic embedding or response is supplied. The
topic-calibrated arm uses the same five-value regularization grid
$C\in\{0.01,0.1,1,10,100\}$. Both retain scalar, graph, one- and two-lag policies
and a persistence control. Evaluation uses 1,000 autonomous draws, with common
random numbers across policies and nested prefixes at $T=3,6$. The macro metric
compares the ensemble-mean surrogate trajectory with each observed LLM trajectory
before averaging replicas, sizes and questions. It is not a distance between two
fully sampled LLM and surrogate path distributions.

\subsection{Technical replay and unmatched question sets}
The sole replay target is the Gemma/u15 global-input test ($N=64$, replica 2),
stopped at round 5/node 20 with a transport error and no response. A separately
frozen full six-round replay produces 384 responses under unchanged model, prompt,
decoding and sampling identities. It is stored separately, never spliced into the
old prefix. This restores one missing case, not an extra test realization, and
does not alter calibration data. Original and amended analyses remain separate.

Qwen/u09's received extra JSON field \texttt{should} remains a schema failure;
that episode is neither repaired nor replayed and its matched panel is excluded.
The final panel therefore has 20 Gemma and 19 Qwen questions. Comparisons
between backbones use different question sets and must not be read as a paired
quality ranking. The paired comparisons within each backend retain identical data
across policy families and both calibration arms.

Table~\ref{tab:topics} reports the graph-versus-scalar contrasts and
Table~\ref{tab:topicabsolute} their absolute-error context. The Gemma sensitivity
in Table~\ref{tab:topicretry} changes the question set from 19 to 20; it does not
isolate an effect of replay on a fixed panel. The original and amended intervals
are descriptive, not multiplicity-adjusted population guarantees.

\begin{appendixtable}
 \caption{New-question graph-versus-scalar reductions at $T=6$, averaged equally
 over $N=16,64$. Positive favors graph. Brackets are descriptive 95\%
 whole-question intervals, without multiplicity adjustment. All entries use the
 technical-replay overlay: 20 Gemma and 19 Qwen questions. Primary comparisons
 are the topic-specific local-input rows; other rows are secondary. $E_m$ abbreviates $E_m^{\mathrm{new}}$ in Eq.~\eqref{eq:mae-new}.}
 \label{tab:topics}\centering\small
\begin{tabular}{lllrr}
\toprule
Backend & Input & Calibration & $\Delta\ell$ & $\Delta E_m$\\
\midrule
Gemma & global & Source-frozen & $0.0260\;[0.0094, 0.0462]$ & $0.0000\;[-0.0015, 0.0017]$\\
Gemma & global & Topic-specific & $-0.0197\;[-0.0315, -0.0085]$ & $-0.0084\;[-0.0180, -0.0001]$\\
Gemma & local & Source-frozen & $-0.0640\;[-0.1287, -0.0062]$ & $-0.0244\;[-0.0757, 0.0290]$\\
Gemma & local & Topic-specific & $0.0099\;[-0.0049, 0.0259]$ & $-0.0056\;[-0.0311, 0.0150]$\\
Qwen & global & Source-frozen & $-0.0350\;[-0.0409, -0.0285]$ & $-0.0704\;[-0.1060, -0.0325]$\\
Qwen & global & Topic-specific & $-0.0578\;[-0.0925, -0.0280]$ & $-0.0224\;[-0.0525, 0.0019]$\\
Qwen & local & Source-frozen & $-0.0202\;[-0.0432, 0.0017]$ & $-0.0639\;[-0.0863, -0.0371]$\\
Qwen & local & Topic-specific & $-0.0181\;[-0.0586, 0.0140]$ & $0.0037\;[-0.0232, 0.0304]$\\
\bottomrule
\end{tabular}

\end{appendixtable}

\begin{appendixtable}
 \caption{Absolute new-question errors on the same overlay. Lower is better;
 ``History'' means the two-lag policy. Source-frozen policies receive no new-topic
 calibration. Topic-specific policies receive designated training and validation
 episodes, never held-out test observations. Full one-lag and persistence outcomes
 are retained in the machine-readable results. $E_m$ abbreviates $E_m^{\mathrm{new}}$ in Eq.~\eqref{eq:mae-new}.}
 \label{tab:topicabsolute}\centering\small
\begin{tabular}{lllrrrrrr}
\toprule
 & & & \multicolumn{3}{c}{One-step log loss} & \multicolumn{3}{c}{Magnetization MAE}\\
Backend & Input & Calibration & Scalar & Graph & History & Scalar & Graph & History\\
\midrule
Gemma & global & Source-frozen & 0.8639 & 0.8379 & 0.8494 & 0.4614 & 0.4614 & 0.4796\\
Gemma & global & Topic-specific & 0.4029 & 0.4226 & 0.3946 & 0.1372 & 0.1456 & 0.1533\\
Gemma & local & Source-frozen & 0.5351 & 0.5990 & 0.5790 & 0.5467 & 0.5711 & 0.5860\\
Gemma & local & Topic-specific & 0.4595 & 0.4496 & 0.4289 & 0.2234 & 0.2290 & 0.2338\\
Qwen & global & Source-frozen & 0.1622 & 0.1973 & 0.2092 & 0.3363 & 0.4067 & 0.4983\\
Qwen & global & Topic-specific & 0.1799 & 0.2377 & 0.2286 & 0.1914 & 0.2137 & 0.2012\\
Qwen & local & Source-frozen & 0.2712 & 0.2914 & 0.2847 & 0.2806 & 0.3445 & 0.3259\\
Qwen & local & Topic-specific & 0.2808 & 0.2989 & 0.2969 & 0.2100 & 0.2063 & 0.2121\\
\bottomrule
\end{tabular}

\end{appendixtable}

\begin{appendixtable}
 \caption{Gemma replay sensitivity: graph-versus-scalar point reductions before
 and after restoring u15. Question composition changes from 19 to 20; these are
 not paired estimates of a replay effect. Qwen is unchanged. Original and amended
 per-question values and uncertainty intervals are archived separately. $E_m$ abbreviates $E_m^{\mathrm{new}}$ in Eq.~\eqref{eq:mae-new}.}
 \label{tab:topicretry}\centering\small
\begin{tabular}{llrrrr}
\toprule
 & & \multicolumn{2}{c}{Original ($q=19$)} & \multicolumn{2}{c}{Replay ($q=20$)}\\
Gemma input & Calibration & $\Delta\ell$ & $\Delta E_m$ & $\Delta\ell$ & $\Delta E_m$\\
\midrule
global & Source-frozen & 0.0271 & 0.0001 & 0.0260 & 0.0000\\
global & Topic-specific & -0.0198 & -0.0078 & -0.0197 & -0.0084\\
local & Source-frozen & -0.0662 & -0.0212 & -0.0640 & -0.0244\\
local & Topic-specific & 0.0093 & -0.0097 & 0.0099 & -0.0056\\
\bottomrule
\end{tabular}

\end{appendixtable}

\Needspace{12\baselineskip}
\section{API transfer: complete results and validation replay}\label{app:api}
\subsection{Generation and calibration}
The request settings are temperature 0.7, a 128-token output cap, reasoning
disabled, and the original strict JSON parser. Source training uses 192 node
transitions per backend; new-topic training uses 96 per question.
The frozen grid has sixteen source episodes and 120 new-topic episodes.
The eight source-policy fits precede all new-topic calls. These model
identifiers name the requested API versions; no parameter count or static
internal model snapshot is inferred from the provider's names.

\subsection{Validation-episode replacement}
The separately frozen technical amendment targets GLM question \texttt{u14},
local input, validation $N=32$, replica 0. The original failures occurred in
round 3 at nodes 23 and 28; neither had a saved response. The full-episode replay
retains the model, prompts, initial conditions and decoding. Its HTTP timeout
increases from 45 to 120 seconds and it uses one worker; the nineteen previously
unstarted episodes retain the original two workers. No automatic call retries
are introduced. All 192 replacement responses pass the unchanged parser.

This is a validation replacement, unlike the test replacements in
Appendices~\ref{app:long} and~\ref{app:topics}. It makes the missing GLM/u14
calibration and matched evaluation possible, but supplies no extra test
trajectory. The failed record and original complete-panel analysis are retained.

Table~\ref{tab:api-original} preserves the primary comparison before adding
the replacement validation trajectory. Table~\ref{tab:api-secondary} gives all
six-round secondary pairwise policy contrasts. Table~\ref{tab:api-absolute}
reports absolute errors, including persistence. Each comparison uses matched
complete question panels, not a different subset for each policy.

Separate OpenRouter technical pilots do not enter any scientific comparison;
their response-format investigations did not amend the Alibaba parser.

\begin{appendixtable}
 \caption{Original complete-panel API analysis, before the GLM validation
 replay. The corresponding amended estimates appear in
 Table~\ref{tab:api-primary}. Both use the same prespecified primary comparison
 and interval convention. The replay restores a missing validation panel;
 it does not create another held-out realization. $E_m$ abbreviates $E_m^{\mathrm{new}}$ in Eq.~\eqref{eq:mae-new}.}
 \label{tab:api-original}
 \centering\small
\begin{tabular}{llrr}
\toprule
Model & $q$ & $\Delta\ell$ & $\Delta E_m$\\
\midrule
GLM 5.2 & 19 & $0.0128\;[-0.0282,0.0431]$ & $0.0468\;[-0.0024,0.1240]$\\
Qwen 3.8 Max & 20 & $0.0959\;[0.0313,0.1524]$ & $-0.0035\;[-0.0466,0.0349]$\\
\bottomrule
\end{tabular}

\end{appendixtable}

\begin{appendixtable}
 \caption{All prespecified secondary six-round API policy contrasts.
 Positive error reductions favor the named alternative. History means
 additional spin/field lags, not the agent's full semantic history.
 Intervals are descriptive whole-question bootstrap intervals and are not
 multiplicity-adjusted. $E_m$ abbreviates $E_m^{\mathrm{new}}$ in Eq.~\eqref{eq:mae-new}.}
 \label{tab:api-secondary}
 \centering\small
\begin{tabular}{lllrr}
\toprule
Model & Calibration & Comparison & $\Delta\ell$ & $\Delta E_m$\\
\midrule
GLM 5.2 & Source-frozen & Graph vs scalar & $0.0326\;[0.0177,0.0490]$ & $0.0101\;[-0.0411,0.0624]$\\
GLM 5.2 & Source-frozen & One lag vs graph & $-0.0536\;[-0.0867,-0.0249]$ & $0.0097\;[-0.0189,0.0387]$\\
GLM 5.2 & Source-frozen & Two lags vs graph & $-0.0572\;[-0.0883,-0.0277]$ & $0.0088\;[-0.0206,0.0376]$\\
GLM 5.2 & Topic-specific & One lag vs graph & $0.0111\;[-0.0030,0.0305]$ & $-0.0071\;[-0.0158,-0.0003]$\\
GLM 5.2 & Topic-specific & Two lags vs graph & $0.0293\;[0.0021,0.0674]$ & $-0.0083\;[-0.0195,0.0012]$\\
Qwen 3.8 Max & Source-frozen & Graph vs scalar & $0.1076\;[0.0696,0.1454]$ & $0.0759\;[0.0034,0.1458]$\\
Qwen 3.8 Max & Source-frozen & One lag vs graph & $0.0062\;[-0.0082,0.0226]$ & $-0.0240\;[-0.0551,0.0091]$\\
Qwen 3.8 Max & Source-frozen & Two lags vs graph & $0.0063\;[-0.0062,0.0192]$ & $-0.0139\;[-0.0381,0.0129]$\\
Qwen 3.8 Max & Topic-specific & One lag vs graph & $0.0234\;[0.0009,0.0491]$ & $0.0007\;[-0.0282,0.0360]$\\
Qwen 3.8 Max & Topic-specific & Two lags vs graph & $0.0390\;[-0.0055,0.0981]$ & $-0.0068\;[-0.0351,0.0213]$\\
\bottomrule
\end{tabular}

\end{appendixtable}

\begin{appendixtable}
 \caption{Absolute API-model errors at $T=6$, local input, held-out $N=16$,
 after the separately audited validation replay. Smaller is better.
 $\ell$ is observed-transition log loss and $E_m$ is autonomous
 mean-magnetization trajectory MAE. All policy families and persistence
 are retained, with descriptive 95\% question-bootstrap intervals. $E_m$ abbreviates $E_m^{\mathrm{new}}$ in Eq.~\eqref{eq:mae-new}.}
 \label{tab:api-absolute}
 \centering\small
\begin{tabular}{lllrr}
\toprule
Model & Calibration & Policy & $\ell$ & $E_m$\\
\midrule
GLM 5.2 & Control & Persistence & $2.0723\;[1.8133,2.3170]$ & $0.5844\;[0.4448,0.7271]$\\
GLM 5.2 & Source-frozen & Graph & $0.2249\;[0.1728,0.2804]$ & $0.5894\;[0.3993,0.7910]$\\
GLM 5.2 & Source-frozen & One lag & $0.2784\;[0.2051,0.3536]$ & $0.5797\;[0.4067,0.7792]$\\
GLM 5.2 & Source-frozen & Two lags & $0.2821\;[0.2084,0.3580]$ & $0.5806\;[0.4068,0.7862]$\\
GLM 5.2 & Source-frozen & Scalar & $0.2575\;[0.2022,0.3129]$ & $0.5995\;[0.4288,0.7791]$\\
GLM 5.2 & Topic-specific & Graph & $0.2441\;[0.1501,0.3488]$ & $0.1638\;[0.1006,0.2366]$\\
GLM 5.2 & Topic-specific & One lag & $0.2331\;[0.1506,0.3421]$ & $0.1709\;[0.1082,0.2390]$\\
GLM 5.2 & Topic-specific & Two lags & $0.2148\;[0.1405,0.2938]$ & $0.1721\;[0.1102,0.2402]$\\
GLM 5.2 & Topic-specific & Scalar & $0.2563\;[0.1745,0.3467]$ & $0.2082\;[0.1267,0.3053]$\\
Qwen 3.8 Max & Control & Persistence & $2.7919\;[2.1587,3.3675]$ & $0.4417\;[0.3198,0.5615]$\\
Qwen 3.8 Max & Source-frozen & Graph & $0.2158\;[0.1718,0.2622]$ & $0.4535\;[0.3218,0.5913]$\\
Qwen 3.8 Max & Source-frozen & One lag & $0.2097\;[0.1787,0.2413]$ & $0.4776\;[0.3508,0.6153]$\\
Qwen 3.8 Max & Source-frozen & Two lags & $0.2095\;[0.1743,0.2435]$ & $0.4674\;[0.3366,0.5979]$\\
Qwen 3.8 Max & Source-frozen & Scalar & $0.3235\;[0.2881,0.3605]$ & $0.5294\;[0.4061,0.6624]$\\
Qwen 3.8 Max & Topic-specific & Graph & $0.2582\;[0.1925,0.3371]$ & $0.2756\;[0.1863,0.4062]$\\
Qwen 3.8 Max & Topic-specific & One lag & $0.2348\;[0.1822,0.2918]$ & $0.2750\;[0.1873,0.3976]$\\
Qwen 3.8 Max & Topic-specific & Two lags & $0.2192\;[0.1800,0.2639]$ & $0.2824\;[0.1964,0.4042]$\\
Qwen 3.8 Max & Topic-specific & Scalar & $0.3541\;[0.2953,0.4161]$ & $0.2721\;[0.1889,0.4001]$\\
\bottomrule
\end{tabular}

\end{appendixtable}

\Needspace{12\baselineskip}
\section{Prompts, personas and statements}\label{app:protocol-inputs}
These are the frozen inputs for our finite-grid and follow-up campaigns, not
the public-data prompts. Section~\ref{sec:fresh}
specifies graphs, initialization and scheduling; Appendices~\ref{app:fresh}--\ref{app:api}
give generation and parsing settings. Objective truth labels below are metadata,
never fitted-policy covariates or instructions about the model's answer.

\subsection{System instruction and input construction}
The system instruction is:
\begin{quote}\small
You are one fictional participant in an opinion simulation. Decide whether you agree with the statement. Opinion 1 means agree; -1 means disagree. A relationship 1 means you generally trust that neighbour; -1 means you generally distrust that neighbour. You may keep or change your opinion. Return only a JSON object with exactly two fields: opinion (integer 1 or -1), and message (a public explanation of at most 25 words). Do not give private reasoning or call tools.
\end{quote}
\begingroup\raggedright
The user message is a JSON object with keys \texttt{statement}, \texttt{persona},
\texttt{your\_previous\_opinion}, and \texttt{your\_previous\_message}.
Global input adds \texttt{population\_mean}; local input instead adds
\texttt{neighbours}, ordered by increasing node index, each with
\texttt{relationship}, \texttt{opinion}, and \texttt{message}.
It is serialized with Python's default \texttt{json.dumps}; no further history is appended.
Initial public messages are ``My initial position is to agree with the statement.''
or ``My initial position is to disagree with the statement.'' according to the imposed spin.
\par\endgroup
With zero-based node index $i$, persona $i\bmod4$ is assigned from this roster:
\begin{enumerate}
\setlength{\itemsep}{1pt}\setlength{\parsep}{0pt}
\item You weigh evidence and practical implications carefully.
\item You assess tradeoffs rather than following a majority automatically.
\item You are skeptical of unsupported claims and willing to change your mind.
\item You consider assumptions and long-term consequences.
\end{enumerate}
\subsection{Original eight statements}
\begin{description}
\setlength{\itemsep}{1pt}\setlength{\parsep}{0pt}
\item[o1, objective, truth +1] The integer 29 is a prime number.
\item[s1, subjective] Cities should replace some car parking spaces with protected bicycle lanes.
\item[o2, objective, truth -1] For an independent fair coin, a sequence of heads makes tails more likely on the next toss.
\item[s2, subjective] A four-day workweek with unchanged weekly pay should become the standard.
\item[o3, objective, truth +1] The interior angles of a triangle in Euclidean geometry sum to 180 degrees.
\item[s3, subjective] Public museums should offer free admission to all visitors.
\item[o4, objective, truth -1] Under standard atmospheric pressure, pure water freezes at 10 degrees Celsius.
\item[s4, subjective] University lectures should be recorded and made available to students by default.
\end{description}
The source-frozen follow-ups retain the four subjective statements s1--s4.
\subsection{Twenty new subjective statements}
\begin{description}
\setlength{\itemsep}{1pt}\setlength{\parsep}{0pt}
\item[u01] Online shops should charge customers for returning non-defective items even if free returns increase sales.
\item[u02] Video review in professional sports should be available only when a team uses a limited challenge.
\item[u03] Orchestras should use anonymous auditions for every performance-based hiring decision.
\item[u04] Residential streets should dim their lighting after midnight to reduce energy use and disturbance.
\item[u05] Public libraries should allocate most of their new-book budget according to reader requests rather than librarian selection.
\item[u06] Domestic cats in towns should be kept indoors rather than allowed to roam freely.
\item[u07] Restaurants should require a non-refundable deposit for reservations during busy periods.
\item[u08] Researchers should be allowed to keep newly collected publicly funded data exclusive for one year before sharing it.
\item[u09] A structurally sound historic building should be preserved even when replacing it would provide substantially more usable space.
\item[u10] Social platforms should show posts in chronological order by default rather than rank them for engagement.
\item[u11] Public disaster assistance should prioritize relocation over rebuilding homes in repeatedly flooded areas.
\item[u12] Secondary schools should require students to keep personal phones in lockers throughout the school day.
\item[u13] Physical shops should be required to accept cash even when electronic payments are cheaper to process.
\item[u14] Zoos should prioritize breeding endangered species over exhibiting animals that attract the most visitors.
\item[u15] Public space-exploration budgets should prioritize robotic missions over sending people into space.
\item[u16] Residential neighborhoods should prohibit routine delivery drones to protect quiet outdoor spaces.
\item[u17] New housing developments should favor shared community gardens over private gardens for individual homes.
\item[u18] New large data centers should be required to supply usable waste heat to nearby buildings when technically feasible.
\item[u19] Competitions should award shared first prizes for tied scores rather than require a tie-breaking round.
\item[u20] Charities should keep a substantial financial reserve even when spending it now could meet urgent needs.
\end{description}
These fixed, nonrandom statements also identify the exclusions and technical
replays in Appendices~\ref{app:long}--\ref{app:api}.

\Needspace{12\baselineskip}
\section{Post-hoc investigation of the history-policy contrast}
\label{app:history-investigation}

\subsection{Common target and source-only controls}
These stages were designed sequentially after inspecting earlier results on the
same local campaigns. They concern previously inspected topics, not an
independent new-domain validation. The main panel contains the 19 questions
complete for both backends (u01--u20 except u09), $N=16,64$, and test replicas
1,2: 152 six-update episodes. Each comparison averages replicas within size,
sizes within question, and questions equally. Surrogate ensembles have $R=1000$
independent draws with the original episode seeds and aligned absolute-round
random numbers. The two scores in Eqs.~\eqref{eq:fair-mse} and~\eqref{eq:fair-crps}
are restricted here to updates 4--6. For either score $S$, define
$G_S=S(\mathrm{graph})-S(\mathrm{history2})$; positive values favor history.

Besides the original three-update source fits, the source-only controls retain
prefix, spaced and full twelve-update fits, and add unpadded predictor schedules
$\{3,4,5\}$, $\{9,10,11\}$ and $\{3,\ldots,11\}$. Features always use complete
prefixes, not time-compressed histories. These controls use only the four old
questions s1--s4, a common twelve-round source-training standardizer, no round
coordinate, and penalized mean log loss plus
$\|w\|^2/(2C_{\rm ref}\,192)$ for standardized slopes $w$ (unpenalized intercept).
The same source-validation-selected $C_{\rm ref}$, 0.1 for Gemma and 1 for Qwen,
is used across schedules and families; target scores select nothing. The full
grid and all schedules are retained in the accompanying outputs. Under observed
prefix initialization, the unpadded nine-update history advantage is resolved
for Gemma but not Qwen; the equal-sized early/late comparison is unresolved for
both in the declared sensitivity. Longer source exposure therefore does not
isolate a unique memory mechanism or guarantee transfer.

\subsection{Observed initialization and feedback allocation}
The autonomous forecast starts at state 0. The observed-prefix forecast retains
the actual states 0--3 and then simulates updates 4--6, keeping initial-spin
features equal to state 0. The first three observed updates are not counted as forecasts.
No later observation enters simulation. Table~\ref{tab:investigation-controls}
compares identical late scoring windows but explicitly different conditioning
information. Gemma's change in $G_S$ from autonomous to observed-prefix
initialization is \InvGemmaWarmChangeMSE{} for fair MSE and
\InvGemmaWarmChangeCRPS{} for CRPS, with the stage's adjusted intervals.

For arithmetic attribution inside a fixed surrogate, let $F_{ab}$ be its expected
next population mean, averaged over ensemble histories. Subscript $a=0$ uses
observed current-state features and $a=1$ simulated ones; $b$ analogously selects
observed or simulated explicit lag features. Initial state, graph and persona
are common. Define
\begin{align}
A_{\rm current}&=\tfrac12[(F_{10}-F_{00})+(F_{11}-F_{01})],\nonumber\\
A_{\rm lag}&=\tfrac12[(F_{01}-F_{00})+(F_{11}-F_{10})],\nonumber\\
\overline m(t+1)-m(t+1)&=[F_{00}-m(t+1)]+A_{\rm current}+A_{\rm lag}
                         +[\overline m(t+1)-F_{11}].
\label{eq:feedback-accounting}
\end{align}
The two orders allocate the nonlinear interaction symmetrically. The identity
is checked at every evaluated update. For original Gemma history-2, the average
lag term falls from $\InvGemmaLagAuto$ autonomously to $\InvGemmaLagWarm$ after
observed-prefix initialization; Qwen's corresponding values are
$\InvQwenLagAuto$ and $\InvQwenLagWarm$. These are signed-error components, not
additive fractions of MSE or causal mediation in the LLM. Combinations of
observed and simulated features may not occur in the training data, and
$F_{00}-m(t+1)$ includes reference-path noise. The
initialization intervention resets the entire history, not only its lag channel.

\begin{appendixtable}
\centering\small
\begin{tabular}{llrrrr}
\toprule
Backend & Forecast variant & Graph $E_{2,\mathrm f}$ & H2 $E_{2,\mathrm f}$ & $G_{E_{2,\mathrm f}}$ & $G_{C_{\mathrm f}}$ \\
\midrule
Gemma & Autonomous & 0.6668 & 0.7328 & -0.0660 & -0.0438 \\
Gemma & Observed prefix & 0.4989 & 0.3891 & +0.1098 & +0.0497 \\
Gemma & Prefix + offset & 0.0276 & 0.0347 & -0.0071 & -0.0114 \\
Gemma & Autonomous + clock cap & 0.6711 & 0.7397 & -0.0687 & -0.0465 \\
Qwen & Autonomous & 0.3740 & 0.3306 & +0.0434 & +0.0369 \\
Qwen & Observed prefix & 0.2429 & 0.2045 & +0.0384 & +0.0411 \\
Qwen & Prefix + offset & 0.2002 & 0.1726 & +0.0275 & +0.0306 \\
Qwen & Autonomous + clock cap & 0.2171 & 0.1885 & +0.0286 & +0.0242 \\
\bottomrule
\end{tabular}

\caption{Exploratory controls on the common 19-topic local panel, original short-source
graph and history-2 (H2) policies, and updates 4--6. Values are point estimates;
$G$ is graph minus H2 error. ``Observed prefix'' conditions on states 0--3;
``prefix + offset'' additionally adapts one scalar intercept to updates 1--3;
``clock cap'' changes only the raw time feature beyond source support.
Conditional and autonomous rows do not have the same information budget.
Full per-question scores and separate-size sensitivities are retained.}
\label{tab:investigation-controls}
\end{appendixtable}

\subsection{Prefix adaptation and simple negative controls}
Let $\eta_{it}$ be a frozen source policy's logit evaluated on the observed
prefix, and $y_{i,t+1}=[s_i(t+1)+1]/2$. Estimate one offset $\delta$ by minimizing
\begin{equation}
\sum_{i=1}^N\sum_{t=0}^{2}\left\{\log[1+\exp(\eta_{it}+\delta)]
          -y_{i,t+1}(\eta_{it}+\delta)\right\}+\tfrac12\delta^2.
\label{eq:prefix-offset}
\end{equation}
The unit penalty is fixed, not tuned on later updates; all source slopes and
scales remain unchanged. This adjusts the intercept using the first three
observed rounds; it is not a forecast from an unchanged source model.
Gemma's original graph fair-MSE reduction is \InvGemmaOffsetGain{}.
Offsets from three fixed cyclic donor questions (shifts 1,6,12 in sorted question
order, same backend/size/replica) instead give mean graph MSE
$\InvGemmaDonorMSE$, versus $\InvGemmaOffsetGraphMSE$ for the matched prefix.
Donors change both topic and episode/graph; they couple questions and receive
descriptive means, not population-significance claims. The mean donor score is
not the score of an ensemble mixture.

Two simple forecasts use the same prefix: persistence at each node's state 3,
and a two-state Markov policy. For current spin $s\in\{-1,+1\}$ the latter uses
$\widehat p(+1\mid s)=(n_{s,+}+1/2)/(n_s+1)$, where $n_s$ counts prefix
transitions starting at $s$ and $n_{s,+}$ those ending at $+1$, pooling nodes and
updates 1--3 within that episode. Its future nodes are drawn independently.
The Markov baseline has Gemma fair MSE/CRPS
$\InvGemmaMarkovMSE/\InvGemmaMarkovCRPS$ and Qwen
$\InvQwenMarkovMSE/\InvQwenMarkovCRPS$. No adjusted Gemma advantage of the
offset graph/history policies over this baseline is resolved. The original Qwen
policies even after offset adaptation remain much worse: history-2 has
$\InvQwenOffsetHistoryMSE/\InvQwenOffsetHistoryCRPS$. On this panel, the gains
over unadapted source policies do not establish an advantage over the simpler
model fitted to the first three observed rounds.
The unpadded-source arm also retains a negative control: offset adaptation
worsens Qwen history-2 CRPS in its declared adjusted comparison.

\subsection{Extrapolation of the surrogate's time coordinate}
For the four unchanged original fits, replace only the raw coordinate $t$ by
$\min(t,2)$. This preserves all training/validation designs and logits, and all
autonomous simulated states through update 3. It does not clip actual histories
or estimate a new coefficient. For Qwen history-2 the late MSE falls by
$\InvQwenClockPercent\%$; the score reductions are \InvQwenClockGainMSE{} and
\InvQwenClockGainCRPS{}. For Qwen, graph and history-2 improve with either initialization,
whereas Gemma does not improve with the same cap. Qwen's residual autonomous
history advantage remains \InvQwenCappedGapMSE{} and \InvQwenCappedGapCRPS{}.
Thus extrapolating the clock contributes to its error and relative history
advantage, but does not explain that advantage completely. This intervenes in
the fitted surrogate's clock, not an internal LLM clock, and does not prove
that a stationary transfer law is optimal.

\subsection{Message-content intervention at fixed spin states}
The same pinned local installations and decoding settings of
Appendix~\ref{app:fresh} generated one new response per context--responder--message
combination. For each source
backend and common question, take the $N=16$, replica-1 local context at $t=3$
for nodes 0--3, covering all four personas: 152 contexts. Both responders see
every context in two forms. Native messages are unchanged; stance-only messages
replace each own/neighbor explanation by ``I agree with the statement.'' or
``I disagree with the statement.'' according to its unchanged spin. Statements,
personas, graph signs, current spins and underlying spin histories are unchanged,
so the spin/field policies cannot distinguish the treatments. The substitution
changes textual content, style, detail and length together.

Context hashes fix seeds and counterbalance treatment order. A shared integer
seed need not couple random draws identically across architectures. All 608
planned responses are valid, with no technical retries or excluded topics;
before/after model digests agree. These are one-step probes, not 608 societies.
Rates average nodes, then source-context backends within question, then questions.

\begin{appendixtable}
\centering\small
\begin{tabular}{lrrr}
\toprule
Responder & Native agreement (\%) & Stance-only agreement (\%) & Difference [adjusted interval], p.p. \\
\midrule
Gemma & 57.895 & 68.421 & $+10.5263\;[+3.9474,+17.1053]$ \\
Qwen & 54.605 & 54.605 & $+0.0000\;[-3.9474,+3.9474]$ \\
\bottomrule
\end{tabular}

\caption{New text-intervention responses on identical contexts. The difference
is stance-only minus native agreement, in percentage points (p.p.). Brackets
are the six-contrast post-hoc sensitivity intervals, not new-domain validation.}
\label{tab:direct-text-probe}
\end{appendixtable}

The Qwen-minus-Gemma agreement-effect difference is \InvTextDifference{}
percentage points. The three flip-probability contrasts (relative to the fixed
current spin) are \InvTextFlipGemma{} for Gemma, \InvTextFlipQwen{} for Qwen,
and \InvTextFlipDifference{} for their difference; all include zero.
Crucially, descriptive source-context strata give a Gemma agreement shift of
$\InvGemmaOwnTextDelta$ percentage points on Gemma-origin contexts and
$\InvGemmaCrossTextDelta$ on Qwen-origin contexts. These strata differ in the
whole source episode, not only author style. The overall effect therefore does
not establish a textual cause of the original penalty on Gemma's own paths.
Changed paired answers also include decoding randomness; they are not counts
of deterministic causal switches.

\subsection{Uncertainty and retained negative results}
Ordinary paired question intervals use 2,000 bootstrap resamples; within-stage
sensitivities use 100,000 resamples and percentile tails $0.025/K$ and
$1-0.025/K$. The declared families have $K=28$ for dynamics (two backends,
two scores, seven fixed history-gap contrasts); $K=32$ for prefix offsets,
$K=32$ for the two simple baselines, and $K=32$ for clock controls. The offset
family crosses two backends, two source arms, two scores and four contrasts;
the baseline family crosses two baselines, two arms, two policy families, two
backends and two scores. The clock family crosses two backends, two starts,
two scores and four contrasts (each policy's gain, change in $G$, capped $G$).
For text, $K=6$ covers two responder effects and their difference for agreement
and flipping. All other panels, windows and source-context strata are descriptive.
The protocol files retain the complete comparison lists and results, including
the unresolved early/late contrast and adverse offset/clock outcomes.
This is not a multiplicity correction for the entire investigation or a
guarantee of family-wise coverage: these sequential post-hoc stages reuse fixed
topics and reference paths. They localize failure modes of specified fitted
simulators, but neither identify all parameters nor isolate intrinsic LLM
memory, architecture or model-size effects. The independently collected
24-statement test in Sec.~\ref{sec:prospective} limits transfer of the earlier
autonomous history contrast and clock-cap result; the controls in this appendix
must not be read as replicated general explanations of those effects.

\Needspace{12\baselineskip}
\section{Prospective replication: protocol, absolute scores and statements}
\label{app:prospective}
\subsection{Collection and information boundaries}
The test reuses the system instruction and four personas in
Appendix~\ref{app:protocol-inputs}, the original graph/initialization construction,
and the latest-own/neighbor-message prompt, with new statement IDs v01--v24.
Both backends receive matched initial graphs and spins. The local artifacts are
\texttt{gemma4:9b} (installed metadata: 8.0B parameters, Q4\_K\_M) and
\texttt{qwen3.8:latest} (27.3B, Q4\_K\_M); full artifact digests match before
and after collection. Decoding fixes temperature 0.7, prediction limit 128,
context limit 4096, thinking disabled, JSON output and no streaming. Names alone
are not treated as model identities. No new provider/API models enter this test.

The collection has a fixed 48-hour request-start limit per backend and a
180-second request timeout. This operational limit is unrelated to the numerical
surrogate's clock-feature cap. A transport failure, truncated generation, invalid
JSON or schema violation permits one identical technical retry per node; a valid
opinion is terminal regardless of its sign. Two failed attempts terminate that
episode without imputation, and three consecutive exhausted episodes pause that
backend. No replacement questions, seeds or whole-episode retries are permitted.
In the completed collection no failure, technical retry or exclusion occurred.
Raw receipts reconstruct every synchronous update and its exact prompt.

There are six autonomous forecasts: graph, history-2, capped graph, capped
history-2, persistence and two-state. Conditional forecasts include those six
and offset graph and offset history-2. The source coefficients, standardizers
and regularization choices are frozen before collection. Autonomous two-state
probabilities use only the original four-question $N=16$ source-training
transitions, not source validation or new responses. Conditional two-state and
offset procedures use only the episode's observed updates 1--3, with the
pseudocounts and penalty specified in Sec.~\ref{sec:prospective} and
Eq.~\eqref{eq:prefix-offset}. Prefix labels cannot enter later generated
trajectories except through the permitted initial history and adaptation.

All 14 forecasts use 1,000 draws per episode and the original deterministic
episode seed with absolute-round common random numbers. The scoring window is
updates 4--6 for both starts. Fair MSE targets the predicted mean and fair CRPS
the per-time marginal distribution, not the complete joint trajectory law;
Monte Carlo-corrected scores are not clipped. A complete question requires all
eight backend/size/replica episodes. All 24 meet this rule, exceeding the fixed
minimum of 20 complete questions for the planned primary interpretation. That
threshold is not a power calculation. The adjusted intervals in
Table~\ref{tab:prospective-primary} use the predeclared 28-comparison family;
2,000-resample nominal intervals are retained separately.

Secondary outputs retain size-specific, last-update and complete six-update
autonomous scores, persistence comparisons and other fixed cap/offset contrasts.
These are descriptive, not additional primary tests. New and earlier topics
are not pooled, and no horizon, policy or comparison is selected using the
new outcomes. A post-collection arithmetic audit independently reconstructs
the 28 contrast vectors and their fixed bootstrap intervals; future-state
perturbations on the first-ID question check that later labels do not enter
forecasts. These checks concern implementation, not empirical generality.

\Needspace{36\baselineskip}
\subsection{All absolute scores on the common late window}
\begin{appendixtable}
\centering\small
\begin{tabular}{lllrr}
\toprule
Backend & Initialization & Forecast & $E_{2,\mathrm f}$ & $C_{\mathrm f}$ \\
\midrule
Gemma & Autonomous & Graph & 0.02822 & 0.08093 \\
Gemma & Autonomous & History-2 & 0.02940 & 0.07310 \\
Gemma & Autonomous & Capped graph & 0.02808 & 0.07984 \\
Gemma & Autonomous & Capped history-2 & 0.02993 & 0.07358 \\
Gemma & Autonomous & Persistence & 0.84942 & 0.90636 \\
Gemma & Autonomous & Two-state & 0.03108 & 0.09400 \\
\midrule
Gemma & Conditional & Graph & 0.02313 & 0.07264 \\
Gemma & Conditional & History-2 & 0.02089 & 0.06000 \\
Gemma & Conditional & Capped graph & 0.02305 & 0.07164 \\
Gemma & Conditional & Capped history-2 & 0.02202 & 0.06155 \\
Gemma & Conditional & Offset graph & 0.01307 & 0.04325 \\
Gemma & Conditional & Offset history-2 & 0.01435 & 0.04545 \\
Gemma & Conditional & Persistence & 0.01548 & 0.05946 \\
Gemma & Conditional & Two-state & 0.02063 & 0.05766 \\
\midrule
Qwen & Autonomous & Graph & 0.26770 & 0.31537 \\
Qwen & Autonomous & History-2 & 0.25649 & 0.30548 \\
Qwen & Autonomous & Capped graph & 0.23273 & 0.28268 \\
Qwen & Autonomous & Capped history-2 & 0.23532 & 0.28621 \\
Qwen & Autonomous & Persistence & 0.41279 & 0.55295 \\
Qwen & Autonomous & Two-state & 0.39029 & 0.46272 \\
\midrule
Qwen & Conditional & Graph & 0.16905 & 0.28112 \\
Qwen & Conditional & History-2 & 0.14905 & 0.26053 \\
Qwen & Conditional & Capped graph & 0.08968 & 0.19291 \\
Qwen & Conditional & Capped history-2 & 0.07755 & 0.17619 \\
Qwen & Conditional & Offset graph & 0.14431 & 0.26070 \\
Qwen & Conditional & Offset history-2 & 0.12731 & 0.24323 \\
Qwen & Conditional & Persistence & 0.02474 & 0.10862 \\
Qwen & Conditional & Two-state & 0.01350 & 0.05784 \\
\bottomrule
\end{tabular}

\caption{All 56 primary-window absolute scores: two backends, 14 forecasts and
two scores, on all 24 questions. Values average updates 4--6, then replicas,
sizes and questions with equal weights. Lower is better. These are descriptive
point estimates, not an additional ranking selected for significance. Autonomous
forecasts observe state 0 only; conditional forecasts observe states 0--3.
Thus the two initialization groups have different information budgets.
Cap and offset denote the fixed time-feature bound and prefix-only intercept
adaptation, respectively; persistence and two-state are specified simple controls.}
\label{tab:prospective-absolute}
\end{appendixtable}

\Needspace{6\baselineskip}
\subsection{Monte Carlo precision of adjusted bootstrap intervals}
\label{app:bootstrap-precision}
The frozen primary analysis uses 100,000 question-bootstrap draws for each
adjusted interval. At tail probability $0.025/28$, only about 89 draws lie below
the target quantile, so a near-zero endpoint can be sensitive to Monte Carlo
quantile estimation. A separate post-collection numerical audit evaluates all
28 unchanged contrast vectors using ten fixed seeds, 2026091500--2026091509,
and 100,000 draws per seed. Seeds and scope were fixed before this audit;
there are no new LLM responses or fits, and no hypothesis or preferred seed is selected.

For Gemma, consider the conditional fair-MSE comparison between the two-state
model and the graph model with an adjusted intercept. The difference is
two-state minus adjusted-graph error. Its original lower endpoint is
$\ProsGemmaBaselineMSELower$. Diagnostic lower endpoints range
from $\BootPrecisionGemmaMin$ to $\BootPrecisionGemmaMax$, and
\BootPrecisionGemmaUnresolved{} of \BootPrecisionSeeds{} intervals include zero.
For Qwen, the autonomous CRPS comparison is two-state minus capped-history
error. The frozen interval includes zero, but
\BootPrecisionQwenPositive{} of the ten diagnostic intervals
are positive. Neither threshold classification is numerically robust.
The remaining 26 classifications are unchanged across these seeds, including
Qwen's conditional history gains and the conditional two-state advantage.

These repetitions check how precisely the bootstrap endpoints were computed.
They do not repeat the LLM experiment or estimate the probability that a
hypothesis is true. Table~\ref{tab:prospective-primary}
retains every original frozen estimate and interval; none is replaced by a
diagnostic interval. The code and all 28 diagnostics are generated separately
by \texttt{bootstrap\_precision.py}.

\Needspace{12\baselineskip}
\subsection{The 24 frozen subjective statements}
The exact strings below were fixed before any response to them was collected.
Normalized exact-string checks exclude overlap with the eight initial, twenty
earlier topic and sixty public-data statements. Thus the statements have no
exact wording matches to those earlier lists. They are not a representative
sample of topics, need not be semantically independent, and may have been seen
during pretraining. Prompts and collection parameters were
not revised after observing responses.
\begingroup\small
\begin{description}
\item[v01] Manufacturers should provide spare parts and repair manuals for household appliances for ten years after sale.
\item[v02] Household water bills should rise sharply for consumption above a basic allowance rather than use a flat price per liter.
\item[v03] Citizen juries advising local government should be selected by lottery rather than filled by volunteers.
\item[v04] Packaged foods should display an environmental-impact grade alongside their nutritional information.
\item[v05] Employers should provide enclosed workspaces instead of open-plan offices even when this requires more floor space.
\item[v06] Public arts grants should favor first-time applicants over established organizations with successful track records.
\item[v07] Popular hiking trails should require advance reservations when visitor numbers threaten to damage the landscape.
\item[v08] Major scientific prizes should recognize entire research teams rather than a small number of individual researchers.
\item[v09] Advertising aimed at children should be prohibited during entertainment programs intended primarily for children.
\item[v10] Organizations should designate one meeting-free working day each week even if coordinating teams becomes harder.
\item[v11] Agricultural support programs should favor growing diverse crop varieties over maximizing the yield of a single variety.
\item[v12] Product warranties should remain valid when an item is resold to a new owner.
\item[v13] Public procurement should favor durable products with lower lifetime costs even when their initial price is higher.
\item[v14] Outdoor festivals should cap attendance to preserve comfort even when additional visitors would increase revenue.
\item[v15] Public broadcasters should reserve a fixed share of programming for regional languages with few remaining speakers.
\item[v16] Employers should evaluate work samples before reading applicants' educational credentials and previous employers.
\item[v17] Online retailers should offer reusable delivery packaging with a refundable deposit as the default option.
\item[v18] Groundwater users should share a binding collective extraction limit rather than negotiate separate voluntary reductions.
\item[v19] Publicly funded observatories should reserve some telescope time for citizen-science projects rather than allocate all time by expert review.
\item[v20] Oversubscribed public recreation classes should allocate places by lottery rather than by how quickly people register.
\item[v21] Landlords should allow trained assistance dogs in buildings that otherwise prohibit pets.
\item[v22] Government agencies should require open file formats for long-term document storage even when proprietary tools are more convenient.
\item[v23] Food wholesalers should donate edible surplus before selling it for animal feed even when donation is more costly.
\item[v24] Local councils should fund tool-lending facilities so residents can borrow infrequently used household equipment.
\end{description}

\endgroup

\clearpage
\twocolumngrid
\bibliography{references}
\end{document}